\documentclass[paper]{JHEP3}
\pdfoutput=1
\usepackage{amsmath,amssymb,amsthm,amscd,graphicx}
\input epsf.sty

\theoremstyle{definition}

\newcommand{\CC}{{\cal C}}

\newcommand{\CL}{{\cal L}}
\newcommand{\CM}{{\cal M}}
\newcommand{\CN}{{\cal N}}
\newcommand{\CO}{{\cal O}}

\newcommand{\CS}{{\cal S}}

\def\IZ{{\mathbb Z}}
\def\IR{{\mathbb R}}
\def\IC{{\mathbb C}}
\def\IP{{\mathbb P}}

\def\IS{{\mathbb S}}

\newcommand{\re}{{\rm e}}
\newcommand{\ri}{{\rm i}}
\newcommand{\rd}{{\rm d}}
\renewcommand{\d}{\partial}

\newcommand{\be}{\begin{equation}}
\newcommand{\ee}{\end{equation}}
\newcommand{\ba}{\begin{aligned}}
\newcommand{\ea}{\end{aligned}}
\newcommand{\ben}{\begin{eqnarray}\displaystyle}
\newcommand{\een}{\end{eqnarray}}

\newcommand{\sectiono}[1]{\section{#1}\setcounter{equation}{0}}

\newdimen\tableauside\tableauside=1.0ex
\newdimen\tableaurule\tableaurule=0.4pt
\newdimen\tableaustep
\def\phantomhrule#1{\hbox{\vbox to0pt{\hrule height\tableaurule width#1\vss}}}
\def\phantomvrule#1{\vbox{\hbox to0pt{\vrule width\tableaurule height#1\hss}}}
\def\sqr{\vbox{%
  \phantomhrule\tableaustep
  \hbox{\phantomvrule\tableaustep\kern\tableaustep\phantomvrule\tableaustep}%
  \hbox{\vbox{\phantomhrule\tableauside}\kern-\tableaurule}}}
\def\squares#1{\hbox{\count0=#1\noindent\loop\sqr
  \advance\count0 by-1 \ifnum\count0>0\repeat}}
\def\tableau#1{\vcenter{\offinterlineskip
  \tableaustep=\tableauside\advance\tableaustep by-\tableaurule
  \kern\normallineskip\hbox
    {\kern\normallineskip\vbox
      {\gettableau#1 0 }%
     \kern\normallineskip\kern\tableaurule}%
  \kern\normallineskip\kern\tableaurule}}
\def\gettableau#1{\ifnum#1=0\let\next=\null\else
\squares{#1}\let\next=\gettableau\fi\next}

\newcommand{\figref}[1]{Fig.~\protect\ref{#1}}

\preprint{Imperial-TP-2011-ND-01}

\title{Nonperturbative aspects of ABJM theory}

\author{
Nadav Drukker$^a$, Marcos Mari\~no$^{b,c}$ and Pavel Putrov$^c$
\\
$^a$The Blackett Laboratory, Imperial College London,\\
\phantom{$^a$}%
Prince Consort Road, London SW7 2AZ, U.K.\\ 
$^b$D\'epartement de Physique Th\'eorique et $^c$Section de Math\'ematiques,\\
\phantom{$^b$}%
Universit\'e de Gen\`eve, Gen\`eve, CH-1211 Switzerland\\
\\
\email{n.drukker@imperial.ac.uk}, \quad
\email{marcos.marino@unige.ch}, \quad
\email{pavel.putrov@unige.ch}
}

\abstract{
Using the matrix model which calculates the exact free energy of ABJM theory on $\IS^3$ 
we study non-perturbative effects in the large $N$ expansion of this model, {\em i.e.}, in the 
genus expansion of type IIA string theory on AdS$_4\times\IC\IP^3$. We propose a general 
prescription to extract spacetime instanton actions from general matrix models, in terms of period integrals of 
the spectral curve, and we use it to determine them explicitly in the ABJM matrix model, as exact functions of the 't Hooft coupling. 
We confirm numerically that these instantons control the asymptotic growth of the genus expansion. 
Furthermore, we find that the dominant instanton action  
at strong coupling determined in this way exactly matches the action of an Euclidean D2-brane instanton wrapping 
$\IR\IP^3$.
}

\begin{document}

\sectiono{Introduction}

String theory is traditionally defined perturbatively as an asymptotic series expansion over world-sheets 
of different genera embedded into a target space. In general, this series has to 
be supplemented by non-perturbative contributions in order to render it well defined at finite coupling. 
The discovery of D-branes, dualities and M-theory in the 90's were hints on how string theories 
can be completed non-perturbatively. Even better, the AdS/CFT correspondence offered an 
exact non-perturbative definition of string theory (at least in certain backgrounds). Since we 
understand field theories much better than string theory, and indeed some are completely well 
defined, we should be able to say the same about the dual string theories.

The partition function of type IIA string theory on AdS$_4\times \IC\IP^3$ and M-theory 
on AdS$_4\times \IS^7/\IZ_k$ is one of the simplest quantities that can be calculated using 
holography. It is given by the partition function of ABJM theory \cite{abjm} on 
$\IS^3$, analytically continued to strong coupling. This is a non-trivial interacting conformal 
field theory, yet this specific quantity (as well as certain Wilson loops) can be calculated 
exactly by localization \cite{kapustin}. Through this procedure, the full three dimensional 
path integral gets reduced to a finite dimensional integral closely related to the matrix model 
describing topological Chern--Simons on the lens space $\IS^3/\IZ_2$ \cite{csmm,akmv}, as it was pointed out in \cite{mp,dt}.

This matrix model was solved in \cite{akmv,hy,hkr} and its solution was applied to calculating the Wilson loops 
of ABJM theory in \cite{mp}. These results were extended in \cite{dmp} in several directions including 
the calculation of the free energy, which at the planar level scales like $N^{3/2}$, where $N$ is the 
rank of each of the gauge groups. The non-planar corrections to the Wilson loops and free energy 
were also calculated. Indeed, for the free energy, techniques developed in topological string theory allow for the automated 
recursive calculation of higher genus contributions which we carried out up to genus 28. At every genus 
the result is a quasi-modular form, which can be expressed in terms of three basic ones, as outlined 
in the next section. The lowest order ones were written explicitly there, writing the higher ones would fill 
several books.

Within these horrendous expressions lies a lot of information, some of which we will present and 
analyze here. As a first result, we extract from the genus expansion 
explicit expressions for the large radius expansion of M-theory on AdS$_4\times \IS^7/\IZ_k$. This expansion 
can be computed, order by order in the radius expansion, for any value of $k$, and it contains both classical and quantum supergravity 
corrections. Then we look at the large order behavior of this genus expansion in order to deduce the structure of its  
asymptotics and extract information about the spacetime instantons of the theory. A similar study was performed 
in the context of non-critical strings in \cite{shenker}. The instanton effects found in this way were instrumental in the 
discovery of D-branes \cite{polchinski} and were later identified as D-branes in non-critical string theory \cite{martinec,akk}. 
 In our case, by defining the string theory in terms of the gauge theory, which 
can be in turn reduced to a matrix model, we are able to recast the search for non-perturbative 
effects in string theory in terms of non-perturbative effects in matrix models, similar to what happened in non-critical 
string theories. These effects arise from tunneling 
of eigenvalues between different saddle points, and their action is proportional to different period 
integrals of the spectral curve encoding the planar limit of the theory. We verify numerically that these period integrals 
indeed govern the large order behavior of the genus expansion in ABJM theory. One interesting result of this investigation is that, in this theory, 
the genus expansion seems to be Borel summable, since the instantons have complex action \cite{blgzj}. However, Borel summability arises 
as an $\alpha'$-correction, and the instanton action is real in the supergravity limit. 

The matrix model instantons obtained in this way should correspond to non-perturbative objects in the AdS dual. 
As we shall show, the action in the matrix model which governs the growth of the higher genus corrections at strong coupling matches the action 
of a D2-brane wrapping an $\IR\IP^3\subset\IC\IP^3$. Recall that the 
expressions in \cite{dmp} included non-perturbative contributions associated to world-sheet 
instantons wrapping $\IC\IP^1\subset\IC\IP^3$. In M-theory both these configurations are 
M2-brane world-volume instantons, with different orientation with respect to the orbifolded 
cycle in $\IS^7$. It is therefore not surprising that such D2-branes 
show up as a non-perturbative effect in the genus expansion.

The organization and highlights of the rest of this manuscript are as follows. In the next section 
we quote some of the needed results from \cite{dmp}. We present the leading large coupling 
behavior of the free energy up to genus 5, which are all polynomials in certain powers of 
the string length (or Planck length in M-theory). A curious fact is that by a simple redefinition 
of the eleven-dimensional Newton constant these polynomials reduce to monomials, such that the full all-genus 
expansion is in integer powers of the (modified) Newton constant.

In Section 3 we extend previous results on instantons in matrix models and related systems \cite{mswone,mswtwo,kmr}. 
We propose that, when the planar theory is governed by special geometry, 
the instanton actions are just given by linear combinations of periods. We then focus on the matrix model of ABJM theory. 
As discussed in detail in \cite{dmp}, the moduli space of the 't Hooft coupling of this theory is governed by special geometry and 
it has three special points (weak coupling, strong coupling, and conifold point). Correspondingly, we find three different 
instantons whose actions depend on the 't Hooft coupling. The leading non-perturbative effect 
at a given point in moduli space is the instanton with the minimal action there, and 
each of the three instantons we find dominates in each of the three different regions. We test numerically that these actions indeed 
control the large genus behavior of the free energies, confirming in this way our general prescription in the case of ABJM theory. 
In Section 4 we concentrate on the strong coupling regime and show that the dominant instanton there matches the D2-brane mentioned above. 
We discuss our results in Section 5. In the two Appendices we list useful results about metrics and Killing spinors in the 
superstring/M-theory AdS background.

\sectiono{The all genus free energy}

\subsection{Genus expansion in the matrix model}

In this section we present and extend the results for the all genus free energy of ABJM theory obtained in \cite{dmp}, and we give an M-theory/string theory interpretation  
for them. We refer to \cite{dmp} for details on this computation. 

We recall that the ABJM theory \cite{abjm} 
is a quiver Chern--Simons--matter theory in three dimensions with gauge group $U(N)_k \times U(N)_{-k}$ and $\CN=6$ supersymmetry. 
The Chern--Simons actions have couplings $k$ and $-k$, respectively, and the theory contains four bosonic fields $C_I$, $I=1, \cdots, 4$, 
in the bifundamental representation of the 
gauge group. The partition function of the (Euclidean) ABJM theory on $\IS^3$ can be computed by a matrix model \cite{kapustin} which turns out to be 
exactly solvable by using techniques of matrix model theory and topological string theory. In particular, its free energy has a $1/N$ expansion 
of the form 
\be
\label{fgs}
F(\lambda,g_s)=\sum_{g=0}^{\infty} g_s^{2g-2} F_g(\lambda).
\ee
In this equation 
\be
g_s ={2 \pi \ri \over k}, \qquad \lambda={N\over k}
\ee
are respectively the coupling constant and the 't Hooft coupling of the ABJM model. 

In order to write explicit expressions for the genus $g$ free energies, it is useful to introduce a parameter $\kappa\in \IC$ (the meaning 
of this parameter is explained in \cite{mp,dmp}). The 't Hooft coupling is related to $\kappa$ by
 \be
 \label{lamkap}
 \lambda(\kappa)={\kappa \over 8 \pi}   {~}_3F_2\left(\frac{1}{2},\frac{1}{2},\frac{1}{2};1,\frac{3}{2};-\frac{\kappa^2
   }{16}\right), 
\ee
and the genus zero free energy is given by 
\be
\label{comf}
 \partial_\lambda F_0={\kappa \over 4} G^{2,3}_{3,3} \left( \begin{array}{ccc} {1\over 2}, & {1\over 2},& {1\over 2} \\ 0, & 0,&-{1\over 2} \end{array} \biggl| -{\kappa^2\over 16}\right)+{ \pi^2 \ri \kappa \over 2} 
  {~}_3F_2\left(\frac{1}{2},\frac{1}{2},\frac{1}{2};1,\frac{3}{2};-\frac{\kappa^2 
   }{16}\right),
\ee
wheere $G^{2,3}_{3,3}$ is a Meijer function. For $g\ge 1$, the free energies can be written in terms of quasi-modular forms of the modular parameter 
\be
\label{tauex}
\tau=\ri  {K'\left({\ri \kappa \over 4}\right)\over K \left({\ri \kappa \over 4}\right)}.
\ee
For $g=1$, one simply has 
\be
F_1=-\log \, \eta (\tau), 
\ee
where $\eta$ is the usual Dedekind eta function. For $g\ge 2$, the $F_g$ can be written in terms of the quasi-modular forms $E_2$ (the standard Eisenstein series), $b$ and $d$, where 
\be
b=\vartheta_2^4(\tau), \qquad d=\vartheta_4^4(\tau),
\end{equation} 
are standard Jacobi theta functions. More precisely, we have the general structure  
\be
F_g(\lambda)={1\over \left( b d^2 \right)^{g-1}} \sum_{k=0}^{3g-3}
E_2^{k}(\tau) p^{(g)}_k(b,d), \qquad g\ge 2, 
\ee
where $p^{(g)}_k(b,d)$ are polynomials in $b, d$ of modular weight $6g-6-2k$. In \cite{dmp} we described a recursive procedure which gives 
all the genus $g$ free energies unambiguously. 

The genus $g$ free energies $F_g(\lambda)$ obtained in this way are exact interpolating functions of the 't Hooft parameter, and they can be studied in various regimes. When $\lambda\rightarrow 0$ they reproduce the perturbation theory of the matrix model around the Gaussian point, and they behave as
\be
F_g(\lambda)=- {B_{2g} \over g (2g-2)} (2 \pi \ri \lambda)^{2-2g}+\CO(\lambda). 
\ee
They can be also studied in the strong coupling regime $\lambda \rightarrow \infty$, or equivalently, at $\kappa \rightarrow \infty$. In this regime it is more convenient to use the shifted variable 
\be
\label{hatl}
\hat \lambda =\lambda -{1\over 24}={\log ^2\kappa\over 2 \pi^2}+\CO( \kappa^{-2}), \qquad \kappa \gg 1. 
\ee
One finds the following structure. For $F_0$ and $F_1$ one has, at strong coupling, 
\be
\ba
F_0&={4 \pi^3 {\sqrt{2}} \over 3} \hat \lambda^{3/2} + \CO\left(\re^{-2 \pi {\sqrt{2 \hat \lambda}}}\right), \\
F_1&={1\over 6} \log \kappa -{1\over 2} \log\left[ {2 \log \kappa \over \pi} \right]+ \CO\left( {1\over  \kappa^2} \right), 
\ea
\ee
while for $g\ge 2$ one has
\be
\label{fgl}
F_g= f_g\left( {1\over \log \, \kappa} \right)+ \CO\left( {1\over  \kappa^2} \right),
\ee
where 
\be
f_g(x)=\sum_{j=0}^g c_j^{(g)}x^{2g-3+j}
\ee
is a polynomial. One finds, for the very first genera, 
\be
\label{fgpons}
\ba
f_2(x)&=\frac{15 x^3-6 x^2+x}{144}, \\
f_3(x)&=\frac{405 x^6 -135 x^5 +18 x^4-x^3}{5184 }, \\
f_4(x)&=\frac{9945 x^9 -3240 x^8 +450 x^7 -32 x^6 +x^5}{82944}, \\
f_5(x)&=\frac{274590 x^{12}-89505 x^{11} +12960 x^{10}-1050 x^9+48 x^8-x^7}{995328},
\ea
\ee
and the leading, strong coupling behavior is given by 
\be
\label{leading}
 F_g(\lambda)\sim \lambda^{{3\over 2}-g}, \qquad \lambda \to \infty, \quad g\ge0.  
\ee

\subsection{Expansion in the type IIA and M-theory duals}

It is possible to translate the all-genus expansion of the matrix model into expansions in the type IIA and the M-theory duals. In the type IIA dual, the genus expansion 
of the matrix model becomes the genus expansion of superstring theory. In M-theory, the genus expansion becomes an expansion in the Planck length (or, equivalently, in 
Newton's constant). In order to translate the matrix model results in a string/M-theory result we need a precise dictionary relating gauge theory quantities to gravity quantities. 
In particular, one has to take into account the anomalous shifts relating the rank of the gauge group $N$ to the Maxwell charge $Q$, which in turn determines the compactification 
radius $L$ \cite{bh, ahho}. The relation is 
\be
Q=N-{1\over 24}\left( k-{1\over k}\right).
\ee
The charge $Q$ determines the compactification radius in M-theory according to 
\be
\left( {L \over \ell_p}\right)^6= 32 \pi^2 Q k. 
\ee
This means that the shifted variable $\hat \lambda$ introduced in (\ref{hatl}) is given, in M-theory variables, by
\be
\label{mdic}
\hat \lambda ={1\over 32 \pi^2 k^2} \left( {L \over \ell_p}\right)^6 \left( 1-{4 \pi^2 \over 3} \left( {\ell_p \over L}\right)^6 \right). 
\ee
When considering the type IIA expansion, we have to trade $k$ for the string coupling constant $g_{\rm st}$, and the Planck length by the string lenght $\ell_s$. 
In the end we find
\be
\label{ssdic}
\ba
k^2&=g_{\rm st}^{-2} \left( {L \over \ell_s}\right)^2,\\
\hat \lambda &={1\over 32 \pi^2 } \left( {L \over \ell_s}\right)^4 \left( 1-{4 \pi^2 g_{\rm st}^2 \over 3} \left( {\ell_s \over L}\right)^6 \right). 
\ea
\ee
The exponentially small corrections (\ref{fgl}) should correspond, in the type IIA superstring, to worldsheet instantons 
wrapping the $\IC \IP^1$ inside $\IC \IP^3$, and in M-theory to membrane instantons \cite{bbs} wrapping the $\IS^3 \subset \IS^7$. In the following we
will drop these nonperturbative corrections, although they can be of course computed to any given order in the exponentiated worldsheet instanton/membrane action. 

Let us first write down the type IIA superstring expansion. Using the dictionary (\ref{ssdic}) we find
\be
\label{freess}
F=-{ g_{\rm st}^{-2} \over 384\pi^2 } \left( {L \over \ell_s}\right)^8 + {3\over 64} \left( {L \over \ell_s}\right)^2 +{1\over 2} \log\left[ 2\pi \left( {\ell_s \over L}\right)^2 \right] 
+ \sum_{g=2}^{\infty}r_g \left(  \left({\ell_s \over L}\right)^2 \right) g_{\rm st}^{2g-2}
\ee
where 
\be
r_g(x)=\sum_{k=3g-4}^{4(g-1)} r_{g,k} x^k, \qquad g\ge 2, 
\ee
is a polynomial. One finds, for the very first genera, 
\be
\ba
r_2(x)=&-\frac{1}{192}\pi ^2 x^4 \left(5120 x^4-576
   x^2+27\right), \\
r_3(x)=&\frac{1}{32} \pi ^4 x^{10} \left(163840
   x^6-15360 x^4+576 x^2-9\right),\\
r_4(x)=&-\frac{1}{576} \pi ^6 x^{16}
   \left(1158676480 x^8-106168320 x^6+4147200 x^4-82944
   x^2+729\right),\\
r_5(x)=&   \frac{1}{32} \pi ^8 x^{22} \bigl(37916508160 x^{10}-3476029440
   x^8+141557760 x^6\\ 
   & \quad -3225600 x^4+41472 x^2-243\bigr).
   \ea
\ee
Notice that \cite{lm} predicts, for general Sasaki-Einstein manifolds in M-theory, a correction for $F_1$ scaling as $\lambda^{1/2}$, like the second term in (\ref{freess}). It would be interesting to see if the precise numerical coefficient also agrees with theirs. 

We can now work out the M-theory expansion. If we use again the dictionary (\ref{mdic}), we see that the M-theory free energy on ${\rm AdS}_4 \times \IS^7/\IZ_k$ has the structure
\be
\label{freeM}
F=-{1\over 384\pi^2 k} \left( {L \over \ell_p}\right)^9+{3\over 64 k} \left( {L \over \ell_p}\right)^3+{1\over 2} \log \left[ 2 \pi k  \left( {\ell_p  \over L}\right)^3\right] \\
+ {1\over k}  \sum_{n=1}^{\infty} p_n(k)  \left( {\ell_p \over L}\right)^{3n},
\ee
where $p_n(k)$ is a {\it polynomial} in $k$ of degree at most $[(n+3)/3]$. At each order $n$ only a finite number of terms in the original genus expansion contribute, and the maximal genus contributing is 
\be
g=\left[ {n+3\over 2}\right]. 
\ee
The polynomials $p_n(k)$ are given, for the first few orders, by
\be
\ba
p_1(k)&=-{9 \pi^2\over 64}, \\
p_2(k)&={3 \pi^2}, \\
p_3(k)&=-{80 \pi^2\over 3} k^2 -{9 \pi^4 \over 32}.
\ea
\ee
Since each coefficient in the series (\ref{freeM}) is a polynomial in $k$, one can compute from the genus expansion in the matrix model the free energy of M-theory in the large radius expansion, at a given order in $(\ell_p/L)^3$, and for {\it any} value of $k$. 

It turns out that the expansion (\ref{freeM}) has a remarkable hidden structure. As we see, the natural parameter in the power series is 
\be
 \left( {\ell_p  \over L}\right)^3
 \ee
as expected in a generic M-theory expansion. However, if we introduce the following ``renormalized" parameter 
\be
\label{renc}
{\widehat \ell_p\over L} ={\ell_p / L\over \left[ 1- 12 \pi^2 \left( \ell_p/L\right)^6 \right] ^{1/6}},
\ee
it turns out that the expansion can be resummed in the following way,  
\be
\label{freeMal}
F=-{1\over 384\pi^2 k} \left({L\over \widehat \ell_p}\right)^9+{1\over 6} \log \left[ 8 \pi^3 k^3 \left({\widehat \ell_p\over L}\right) ^9 \right] \\
+ \sum_{n=1}^{\infty} d_{n+1} \pi^{2n} k^n  \left({\widehat \ell_p\over L}\right) ^{9n},
\ee
where the coefficients $d_n$ are just rational numbers:
\be
\label{djs}
d_2=-{80\over 3} , \qquad d_3=5120, \qquad d_4=-{18104320 \over 9}, \qquad d_5=1184890880, \qquad \cdots
\ee
This resummation is based on a highly non-trivial property of the polynomials (\ref{fgpons}) which is not at all manifest from their 
matrix model origin, and is begging for an interpretation in the context of M-theory/string theory. A similar simplification can be obtained in the type IIA expansion by introducing a ``renormalized" parameter $\ell_s/L$, which depends also on $g_{\rm st}$. 

What is the interpretation of the M-theory expansion (\ref{freeM}) and its resummation (\ref{freeMal})? In other M-theory expansions (like the two-graviton potential in M(atrix) theory), the terms which go like $(\ell_p/L)^9$ are interpreted like classical supergravity interactions, since they correspond to integral powers of the eleven-dimensional 
Newton's coupling constant. The other terms, with powers which are not multiples of $9$, are usually interpreted as ``quantum gravity" corrections (see for 
example the discussion in \cite{taylor}, IV.A.5). The resummation (\ref{freeMal}) suggest that in this case these quantum gravity corrections can be rewritten in terms of a classical expansion, but involving the ``renormalized" coupling (\ref{renc}).

\sectiono{Instantons and the genus expansion}

\subsection{Instantons in matrix models}

In this subsection we review some results on instantons in matrix models, following the work of \cite{mswone,mswtwo,kmr}, which contains much more details 
and references. 

\FIGURE[ht]{
\includegraphics[height=3cm]{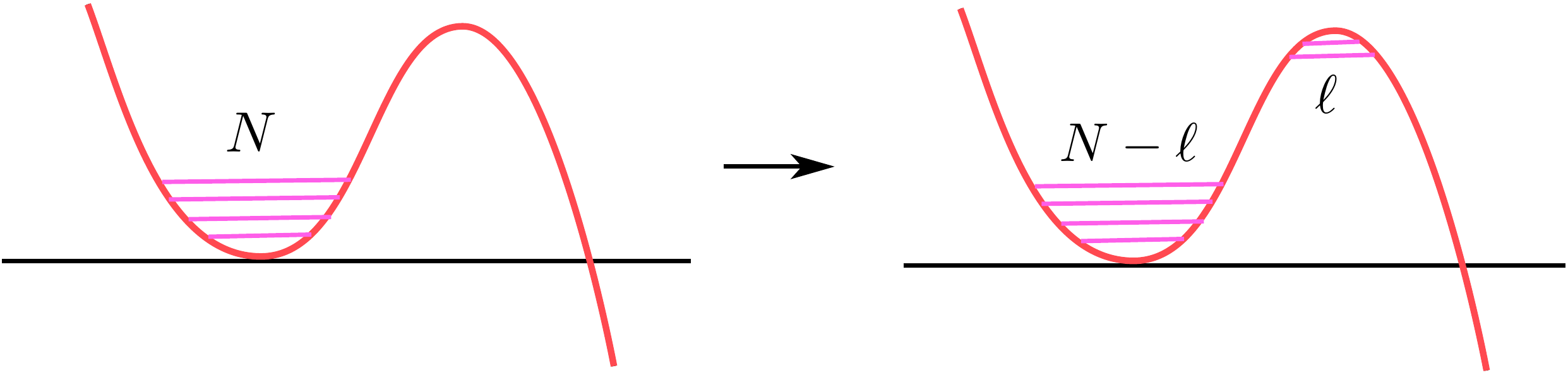} 
\caption{An $\ell$-instanton can be obtained by tunneling $\ell$ eigenvalues from one critical point to another one.} 
\label{tunneling}
}
The study of instantons in matrix models has been pursued in many works, starting with the pioneering papers of David \cite{david}. An important insight, first 
developed in relation to matrix models of two-dimensional gravity, is that instantons are obtained by {\it eigenvalue tunneling}. In order to make the discussion simpler, 
let us consider the cubic matrix model, where the effective potential has two critical points. In the one-cut phase of this model, all eigenvalues are located in the neighborhood of one critical point. The $\ell$-instanton configuration in this phase is simply obtained by removing $\ell$ eigenvalues from this cut and tunneling them to the other critical point, as shown in \figref{tunneling}.  The instanton action for the one-cut phase admits a beautiful geometric interpretation in terms of the spectral curve $y(x)$ describing the planar limit of the matrix model, and it is given by the integral
\be
\label{onecutia}
A_B=\int_a^{x_0} y(x) \rd x, 
\ee
where $a$ is the endpoint of the filled cut, and $x_0$ is the location of the critical point which corresponds to an empty cut ($x_0$ is actually a singular point where the 
spectral curve has a nodal singularity). More geometrically, we can write this as a period integral of the natural meromorphic form $y(x) \rd x$, corresponding to a 
$B$ cycle which goes from the filled cut to the pinched point \cite{ss,kk}:
\be
A_B={1\over 2} \oint_B y(x) \rd x.
\ee
In \figref{curves} (left) we show the pinched curve, where the $A_1$ cycle corresponds to the filled cut, and the $B$ cycle goes from $A_1$ to the pinched cycle. This picture 
extends to one-cut matrix models with generic polynomial potentials: instantons are given by eigenvalue tunneling, and their actions are $B$-type period, going from the 
filled cut to other critical points. 

\FIGURE[ht]{\label{curves}
\includegraphics[height=4cm]{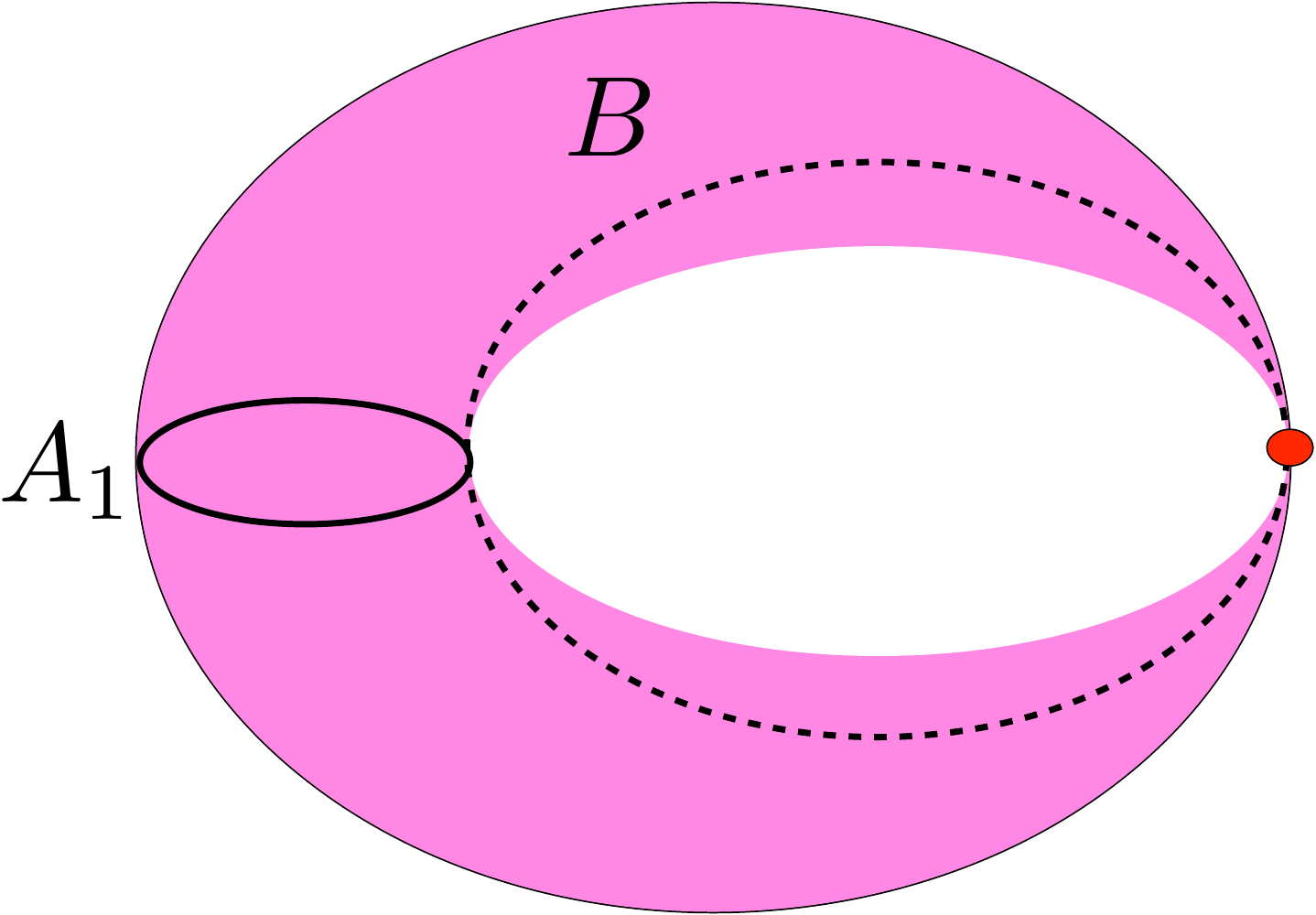} \qquad \qquad
 \includegraphics[height=4cm]{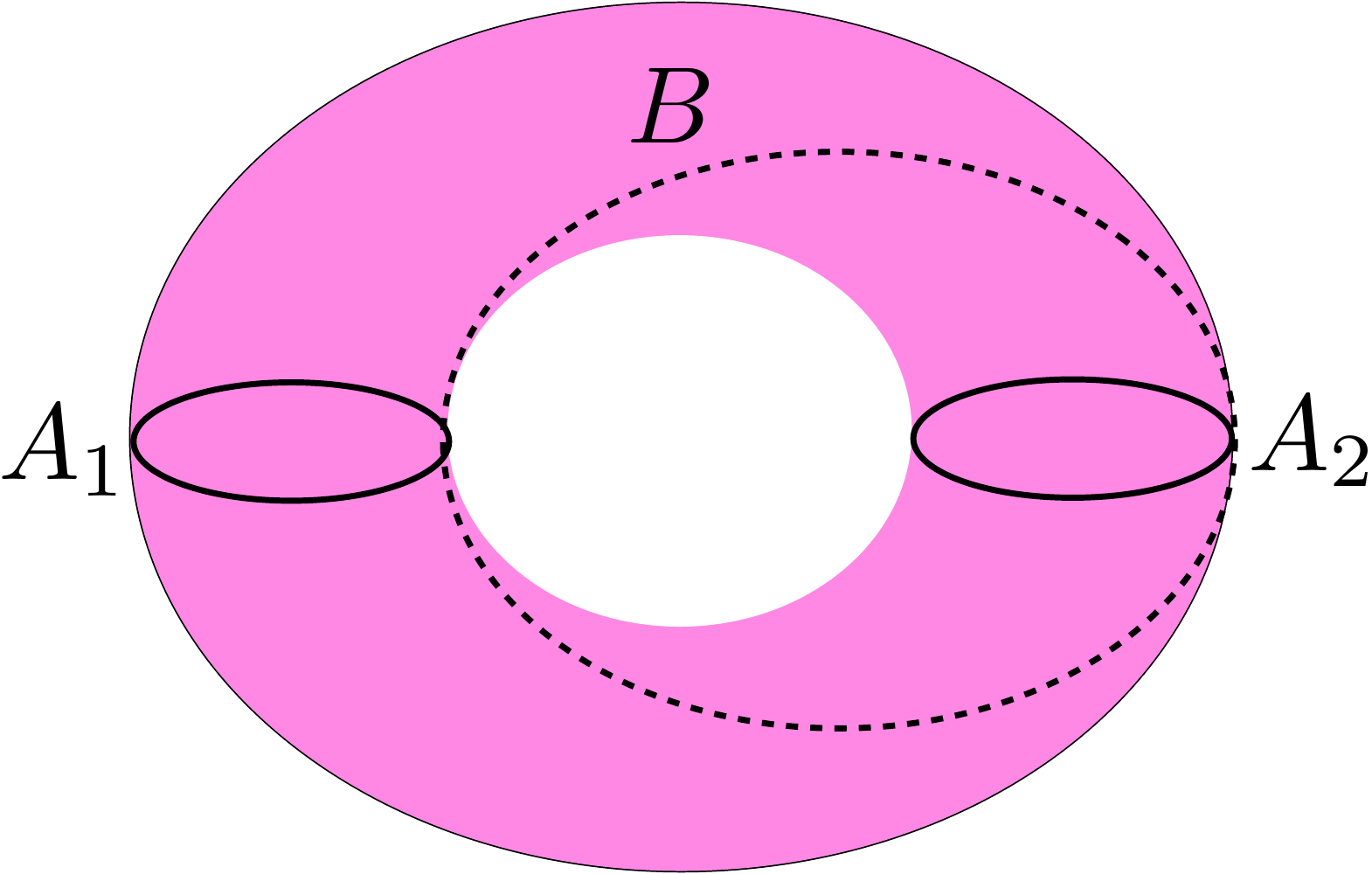} 
\caption{The left hand side shows the spectral curve in the one-cut phase of the cubic matrix model. The instanton action relevant in the double-scaling limit is obtained by 
calculating the $B$-period of the one-form $y(x)\rd x$, which goes from the filled cut $A_1$ to the pinched point. The two-cut phase, in which the pinched point becomes a filled interval, is shown on the right hand side. The instanton action is still given by the $B$-period integral.}}

Based on the connection between instantons and the large order behavior of perturbation theory \cite{lgzj}, we should expect these instantons to control the behavior of the 
genus $g$ amplitudes $F_g$ of one-cut matrix models at large $g$. Indeed, one can verify in examples \cite{mswone} that
\be
\label{asymfg}
 F_g(t) \sim (2g)! (A(t))^{-2g}, \qquad g\gg 1, 
\ee
where $A(t)$ is a period of $y(x) \rd x$. Notice that (\ref{asymfg}) is just the leading behavior of the full asymptotics at large $g$, which involves 
a series of corrections in $1/g$ (see for example \cite{mswone} for more details). The relevant period $A(t)$ 
appearing in the leading asymptotics (\ref{asymfg}) depends on the value of $t$. For small $t$, the behavior of the free energy is dominated by 
its Gaussian part, 
\be
F_g(t) \approx {B_{2g} \over 2g (2g-2) t^{2g}}, \qquad t \rightarrow 0, 
\ee
and the action $A(t)$ in (\ref{asymfg}) is in fact the $A_1$-period going around the filled cut, which is just proportional to the 't Hooft parameter: 
\be
\label{gaussper}
A_{A_1} (t) =2\pi \ri t ={1\over 2} \oint_{A_1} y(x) \rd x. 
\ee
Notice that this period vanishes at the origin $t=0$. In other regions of the $t$-plane, the large genus behavior will be controlled by $B$-periods $A_B(t)$ of the form (\ref{onecutia}). In general, the action controlling the large order behavior at a given point $t$ is the smallest period (in absolute value). Notice that the B-type periods $A_B(t)$ vanish
at critical values of the 't Hooft parameter and the other couplings, so in both cases the instanton action is given by a vanishing cycle in moduli space. 
An equivalent way of formulating the r\^ole of instantons is that their actions give the location of the singularities for the Borel transform of the asymptotic series (\ref{fgs}).

This result can be generalized to the two-cut phase of the cubic matrix model, where the pinched point is now resolved into a second cut $A_2$. 
The instanton action is still given by the $B$-cycle integral, now going from the first cycle $A_1$ to the second cycle $A_2$, and it 
controls the large order behavior of $F_g(t_1, t_2)$ in the appropriate regions of moduli space \cite{kmr}. 
The above analysis of instanton configurations seems to apply to matrix models with generic polynomial potentials. However, there are important matrix models, 
like the Chern--Simons matrix model \cite{csmm}, 
which display a more subtle structure. It was found in \cite{ps}, for example, that due to the multivaluedness of the effective potential, 
the instanton actions in the Chern--Simons matrix model are given by 
\be 
2 \pi \ri  \left( t+ 2\pi \ri n \right), \qquad n \in \IZ. 
\ee
For $n=0$ one recovers the action governing the Gaussian behavior. The instantons with $n=\pm 1$ can be detected through the large order behavior of the genus $g$ free energies, once the Gaussian part is subtracted. 

It is then natural to ask if there is a common structure describing the instantons of general matrix models. All the models we have in mind 
are characterized by the fact that their planar limit is described by special geometry on a local Calabi--Yau manifold, and it is then desirable to describe their 
instantons in that language as well. This is precisely 
what we will do now.

\subsection{Instantons and special geometry}

We will suppose that we are given a local Calabi--Yau manifold, whose geometry is encoded in a spectral curve $y(x)$. This curve can be an algebraic 
curve in $\IC \times \IC$, like the curves arising in polynomial matrix models, or a curve in $\IC^*\times \IC^*$, like the ones arising in Chern--Simons matrix models and in the 
mirrors of toric Calabi--Yau threefolds. We will denote by $\CM$ the moduli space associated to the geometry described by $y(x)$. In order to  
write down the genus $g$ amplitudes $F_g$, one has to choose first a symplectic frame. In order to make this choice manifest we will write $F_g^{(f)}$, where $f$ specifies the choice of frame. The different $F_g^{(f)}$ are related by symplectic transformations and they transform as quasi-modular forms \cite{abk}. 

 Usually, $\CM$ has special points corresponding to physical singularities, and near each of these points there are preferred frames. A famous example is Seiberg--Witten theory \cite{sw}, where $\CM$ is the moduli space of the Seiberg--Witten elliptic curve and there are three special points corresponding to the semiclassical regime, the monopole point and the dyon point. The corresponding frames are usually called electric, magnetic and dyonic, respectively. In the most relevant example for this paper, the ABJM matrix model, the moduli space $\CM$ has three critical points usually called large radius, orbifold and conifold points (see \cite{dmp} for a detailed explanation), so there will be three preferred frames.
  
Our main proposal, based on the results reviewed above, is that instanton actions are always given by complex linear combinations of the periods of special geometry. More precisely, we propose
\be
\label{genia}
A^{(f)}\left(t_i; a_i, b_i,c\right)=\sum_{i=1}^s \left( a_i t^{(f)}_i + b_i {\partial F^{(f)}_0 \over \partial t_i} \right) + c,
\ee
where $s={\rm dim}(\CM)$ and $a_i, b_i, c$ are complex numbers. The first term is the sum gives the contribution of the $A$-cyles, while the second term gives the 
contribution of the $B$-cycles. Notice that this is also the structure of central charges in special geometry. In particular, 
we propose that the large genus behavior of the $F_g^{(f)}$ at generic points of the moduli space is governed by an instanton action of this form. A particular 
r\^ole is played by the instanton actions which govern the large order behavior of the $F_g^{(f)}$ near the singular points of moduli space. We will denote these actions by 
\be
\label{singia}
A^{(f)}_p (t_i)
\ee
where $p$ labels the singular points in $\CM$. According to our proposal, these actions are given by (\ref{genia}), for a specific choice of the constants 
$a_i, b_i, c$ which depends on the point $p$. 

Of course, the main problem 
is to determine the values of the constants $a_i, b_i, c$ which describe the possible instantons in the problem at hand. Unfortunately we don't have a general principle to 
do this. However, when the singular point $p$ is in the interior of $\CM$ (i.e. for singular points of the conifold or orbifold type), we expect the $A^{(f)}_g$ to be given by {\it vanishing periods}. One way to motivate this is to notice that, near the singular points, the $F^{(f)}_g(t_i)$ diverge for all $g$. The instanton action controlling their large order behavior should then vanish at those points. The identification of vanishing periods makes it possible to fix the constants $a_i$, $b_i$ and $c$ in many situations and leads to a determination of the large genus behavior near orbifold and conifold points. At generic points there will be a competition between the different instanton actions, 
and the dominant contribution to the large order behavior will be given by the instanton action which is smaller in absolute value (or, equivalently, by the instanton action which is closest to the origin in the Borel plane). 

Of course, the proposal above recovers and generalizes the known description of instantons in matrix models, where the instanton actions are given by $A$ or $B$ periods, as we have already discussed. In the remaining of this section we will analyze in detail the ABJM matrix model following these general principles, and we will present ample 
evidence that in this model the relevant instanton actions describing the large genus behavior are indeed of the form (\ref{genia}). 


\subsection{Instantons in the ABJM model}

\FIGURE[ht]{\label{moduli}
\includegraphics[height=5cm]{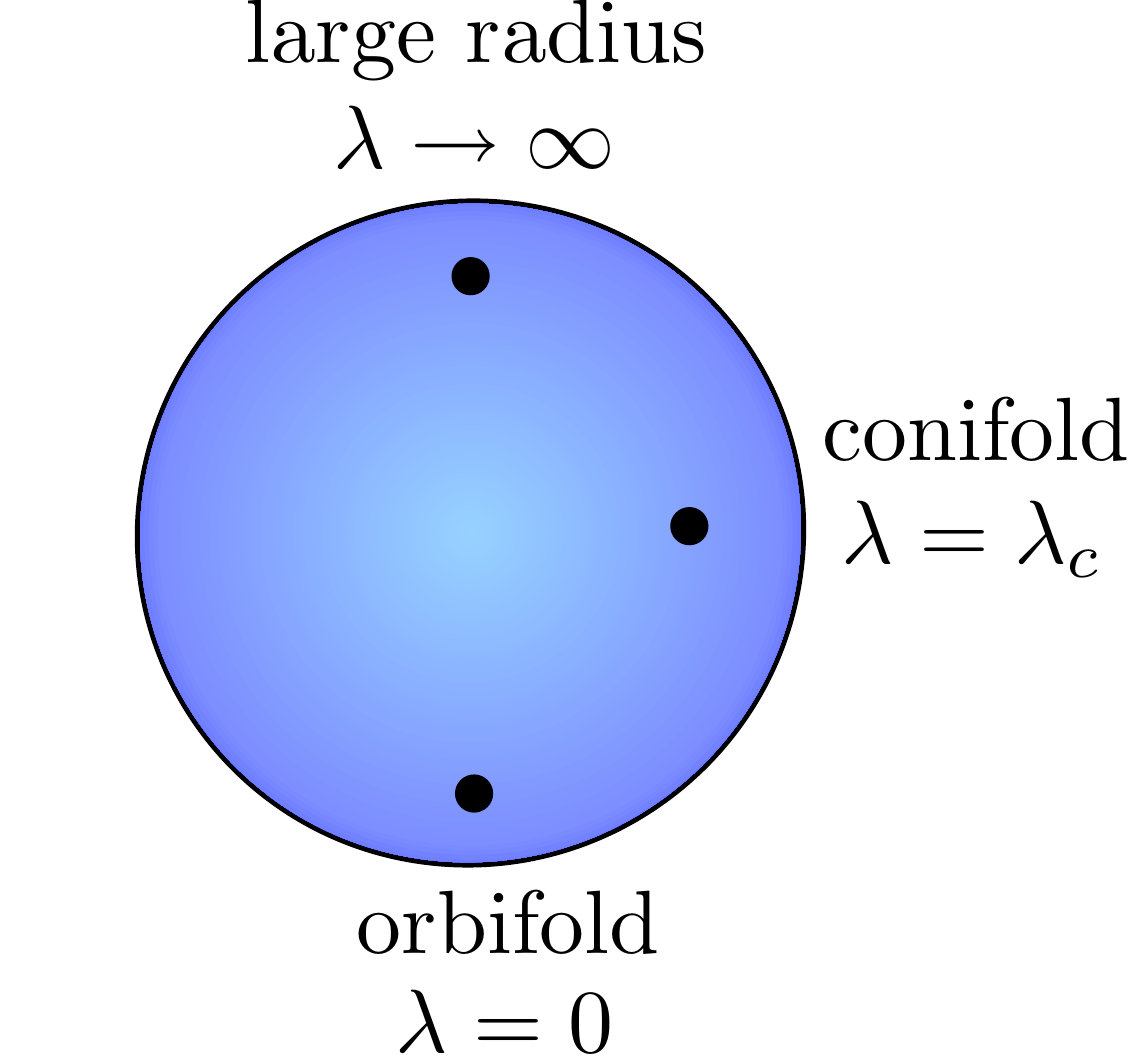} 
\caption{The moduli space of the ABJM theory has three special points.} 
}
The moduli space $\CM$ of the ABJM matrix model was studied in detail in \cite{dmp}, and it is shown schematically in \figref{moduli}. It can be parametrized by $\lambda$ (which is a period), or equivalently by the global modulus $\kappa$. Notice that, although in the original ABJM theory $\lambda$ is a rational number, in the planar solution it is naturally promoted to a complex variable, and $\CM$ will be regarded here as a complex  one-dimensional space. There are three singular points in this moduli space: the orbifold, large radius and conifold points. As explained in \cite{dmp}, the first two points correspond respectively to the weak coupling limit and the strong coupling limits of the ABJM theory. The conifold point, which occurs for $\kappa=-4 \ri$, or equivalently, at
\be
\label{critpoint}
\lambda_c=-{2 \ri K \over \pi^2},
\ee
where $K$ is Catalan's constant, 
has no obvious interpretation in the gauge theory (although we will comment on this later on). 

The frame in which the genus $g$ free energies 
$F_g$ give the $1/N$ expansion of the matrix model is the orbifold or weak coupling frame, as first discovered in \cite{akmv}. We will study this frame first, and we will 
determine the relevant instanton actions near the singular points. In this frame, which we will denote by $w$, the appropriate period coordinate is $\lambda$, 
and the orbifold singularity occurs at $\lambda=0$. 
Near this singularity the relevant instanton action is simply 
\be
\label{wca}
A^{(w)}_{w}(\kappa)=-4 \pi^2 \,  \lambda (\kappa).  
\ee
Since the 't Hooft coupling in the matrix model is $t=2 \pi \ri \lambda$, this action is the standard one (\ref{gaussper}) controlling the Gaussian point. It vanishes of course at $\lambda=0$. On the other hand, it is easy to find the vanishing period at the conifold point, and this leads to the instanton action
\be
\label{wiac}
A^{(w)}_{c}(\kappa)= {\ri \over \pi} {\partial F^{(w)}_0 \over \partial \lambda} +4 \pi^2 \lambda - \pi^2  \\
=\frac{\ri \kappa}{4\pi} G^{2,3}_{3,3} \left(\left. 
  \begin{array}{ccc} 
    \frac{1}{2}, & \frac{1}{2},& \frac{1}{2} \\ 
    0, & 0,&-\frac{1}{2} 
  \end{array} \right| -\frac{\kappa^2}{16}\right)- \pi^2.
\end{equation} 
Finally, we have to consider the large radius, or strong coupling, point. The relevant action turns out to be
\be
\label{stronga}
\ba
A^{({w})}_{s}(\kappa)=&\frac{\ri}{\pi}\frac{\d F^{(w)}_0}{\d\lambda}-\pi^2\\
=&\frac{\ri \kappa}{4\pi} G^{2,3}_{3,3} \left(\left. 
  \begin{array}{ccc} 
    \frac{1}{2}, & \frac{1}{2},& \frac{1}{2} \\ 
    0, & 0,&-\frac{1}{2} 
  \end{array} \right| -\frac{\kappa^2}{16}\right)-\frac{ \pi \kappa}{2} 
  {~}_3F_2\left(\frac{1}{2},\frac{1}{2},\frac{1}{2};1,\frac{3}{2};-\frac{\kappa^2
   }{16}\right)
- \pi^2.
\ea
\end{equation} 
For this action, the coefficients appearing in (\ref{genia}) cannot be determined by requiring it to be a vanishing period, but it has a simple structure, since it is just given by
\be
A^{({w})}_{s}(\kappa)=A^{({w})}_{w}(\kappa)+A^{({w})}_{c}(\kappa). 
\ee
One can verify numerically that it is the right action in the sense that it controls the large order behavior of $F_g^{({w})}$ in the region where 
it dominates the asymptotics, as we will 
show in the next subsection. It is tempting to conjecture that all instanton actions appearing in the theory are just integer linear combinations of $A^{({w})}_{w}(\kappa)$ and $A^{({w})}_{c}(\kappa)$. This is in fact what we would expect if these instantons could be identified with Euclidean D-branes of the string dual, as we will argue 
later.   

\begin{figure}
\centering
\includegraphics[scale=0.8]{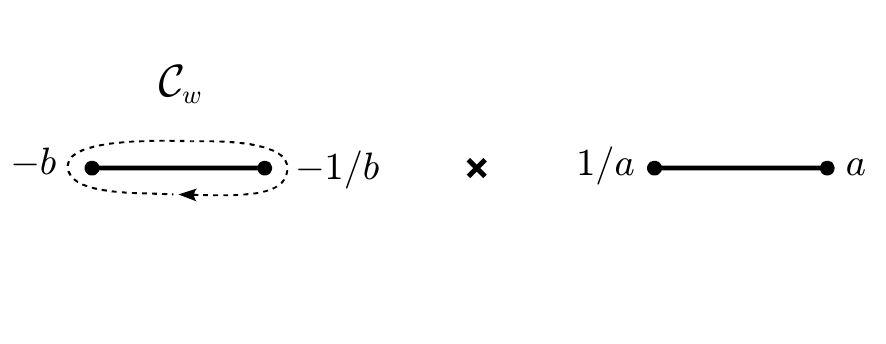}

\includegraphics[scale=0.8]{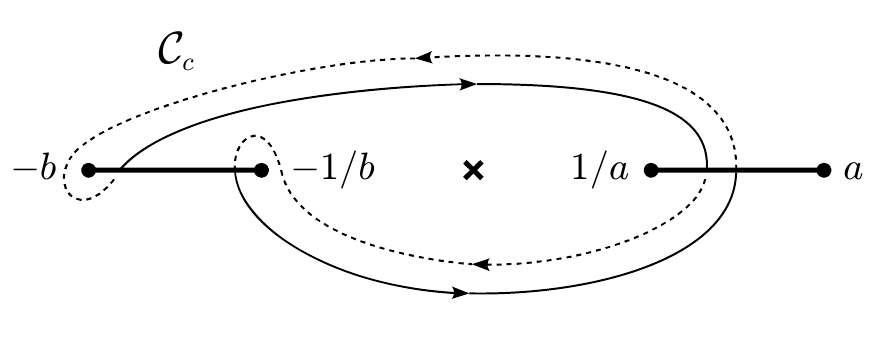} 
\includegraphics[scale=0.8]{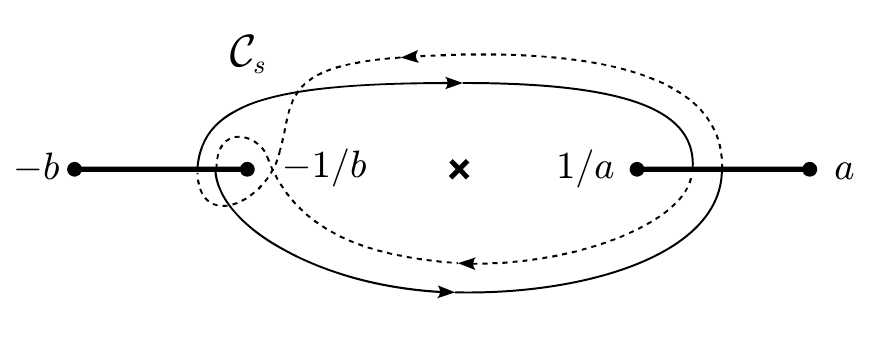} 
\caption{The topology of the contours $\mathcal{C}_w$, $\mathcal{C}_c$, $\mathcal{C}_s$ for a vicinity of the orbifold point in the moduli space.}
\label{contours_fig}
\end{figure} 

The actions (\ref{wca}), (\ref{wiac}) and (\ref{stronga}) can be written as period integrals on the spectral curve of the ABJM matrix model. 
In terms of the variables 
\be
Y=\re^y,\;X=\re^x
\ee
the spectral curve is given by the equation \cite{akmv,hy,mp,dmp}
\begin{equation}
 Y+\frac{X^2}{Y}-X^2+\ri\kappa\,X-1=0\,. \label{curve_eq}
\end{equation}
The Riemann surface of (\ref{curve_eq}) can be represented by two $X$-planes glued along the cuts $[1/a,a]$ and $[-b,-1/b]$. The position of the endpoints can be determined from
\begin{equation}
 a+\frac{1}{a}+b+\frac{1}{b}=4,\qquad 
 a+\frac{1}{a}-b-\frac{1}{b}=2\ri\kappa\,.
\end{equation}
Let us note that $a,\,b\rightarrow 1$ at the orbifold (weak coupling) point, and that $y$ has a logarithmic singularity at the origin (and at infinity) on one of the two $X$-sheets. The actions describing the large $g$ behavior can be represented as
\begin{equation}
 A_p=\frac{1}{2}\oint_{\CC_p}y(x) \rd x
\end{equation} 
where the contours $\mathcal{C}_p$ are depicted in Fig.~\ref{contours_fig}.

As a last remark, notice that these actions appear in pairs $A^{(w)}_p$, $-A^{(w)}_p$, and this is reflected in the fact that the genus expansion that they govern involves only 
even powers of the string coupling constant. Equivalently, there are singularities in the Borel plane of the $g_s$ coupling constant at the points $\pm A^{(w)}_p$. This is also the case in simpler cases related to noncritical string theory, like the Painlev\'e I equation (see for example \cite{gikm}). 
 
\begin{figure}[!ht]
\leavevmode
\begin{center}
\includegraphics[height=4cm]{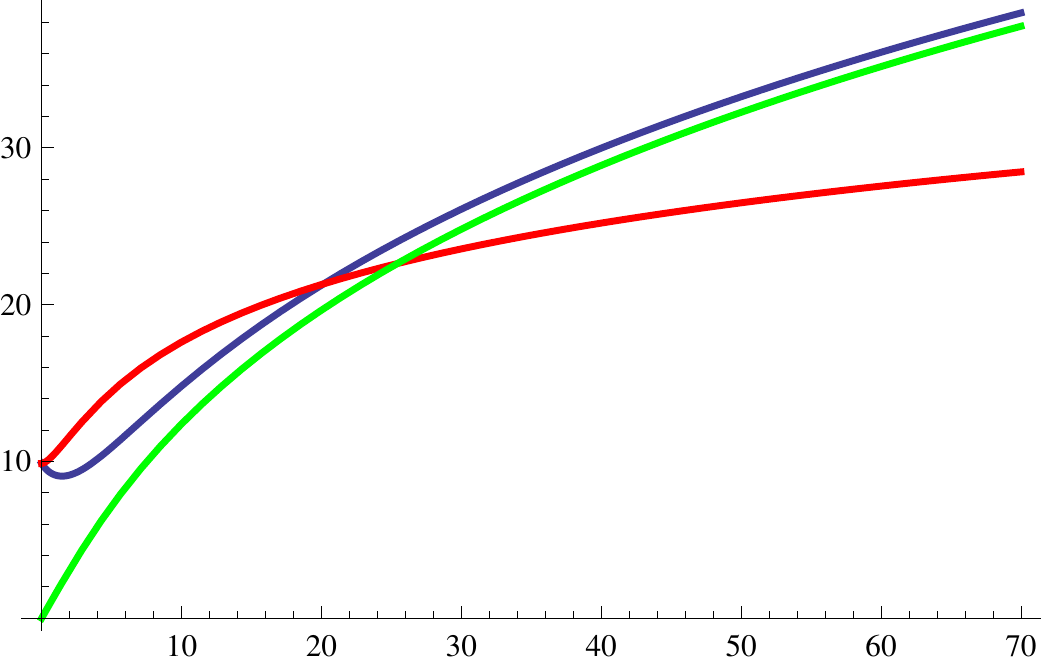}\qquad \qquad 
\includegraphics[height=4cm]{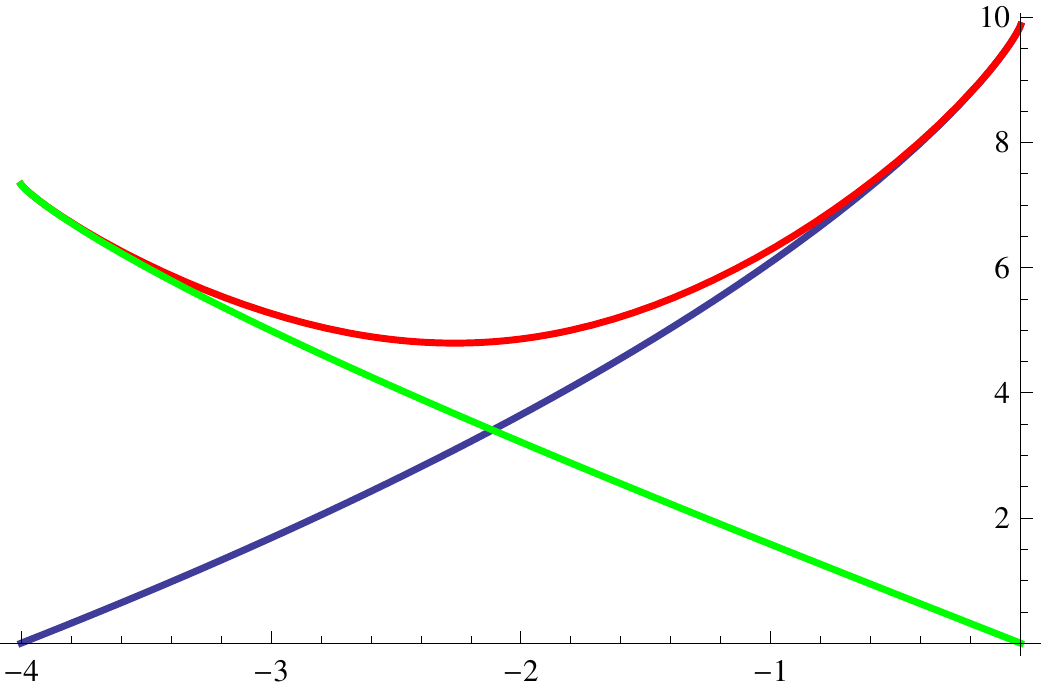}
\end{center}
\caption{In this figure we depict the absolute value of the three instanton actions in the orbifold or weakly coupled frame. On the left side, the horizontal axis represents the positive real axis of the $\kappa$ variable. The curve in green, which vanishes at the origin, is $|A^{(w)}_{w}(\kappa)|$, while the blue and red lines represent 
$|A^{(w)}_{c}(\kappa)|$ and $|A^{(w)}_{s}(\kappa)|$, respectively. Notice that, when $\kappa$ is large (i.e. the strong coupling region), the smallest action in absolute value is 
 $A^{(w)}_{s}(\kappa)$. On the right side, the horizontal axis represents the imaginary axis of the $\kappa$ variable. The conifold action $A^{(w)}_{c}(\kappa)$ vanishes at $\kappa_c=-4\ri$, and therefore dominates the large order behavior near that point. 
}
\label{weakactions}
\end{figure} 

As explained above, at each point in moduli space we expect the large order behavior to be dominated by the smallest action in absolute value. 
In \figref{weakactions} we show 
the absolute value of the instanton actions in the weakly coupled frame, and along two different directions in the complex moduli space parametrized by $\kappa$: the real axis (left) and the imaginary axis (right). For real $\kappa\gg 1$ (which corresponds to the strong coupling regime $\lambda \gg 1$), the smallest instanton action is $A^{({w})}_{s}$, while near the origin the smallest action is $A^{({w})}_{w}$. 
For $\lambda$ imaginary and near the conifold point $\lambda_c$, the smallest instanton action is clearly $A^{({w})}_{c}$. For generic points in the moduli space there is 
a competition between the different actions. For example, for imaginary $\kappa$, there is a point $\kappa_*$ where 
\be
\label{crossingpoint}
\left|A_c^{(w)}(\kappa_*)\right|=\left|A_w^{(w)}(\kappa_*)\right|.
\ee
This is the point where the two lines cross in the graphic on the right side of \figref{weakactions}. 
For $|\kappa|>|\kappa_*|$ we should expect the large order behavior to be controlled by the conifold action 
$A_c^{(w)}$, while for $|\kappa|<|\kappa_*|$ it should be controlled by the weak coupling action $A_w^{(w)}$. We will present explicit checks of these expectations in a moment. 

So far we have made the analysis in the weakly coupled frame, but we can do the analysis in the other preferred frames. It turns out that the relevant instanton 
actions near the singular points are just given by the analytic continuations of the instanton actions in other frames. This is not surprising, since for example vanishing periods 
near a singular point are uniquely defined, independently of the frame. This in particular means that the large genus behavior of the $F_g^{(f)}$ in different frames will be governed by the same instanton action. 

Let us consider for example the conifold frame. An appropriate flat coordinate in this frame is given by \cite{hkr,dmp}
\be
\lambda^{(c)}= {1\over 4 \pi} \sum_{n=0}^\infty\frac{a_n}{(n+1)\,2^{4n}}y^{n+1}, 
\ee
where
\be
y=1+{\kappa^2 \over 16}, \qquad a_n={1\over {2n \choose n}} \sum_{k=0}^n{2k \choose k}{4k \choose 2k}{2n-2k \choose n-k}{4n-4k \choose 2n-2k}.
\ee
This coordinate vanishes at the conifold point $y=0$. The conifold free energies near this point behave as \cite{hkr}
\be
F_g^{({c})} \sim {B_{2g} \over 2g (2g-2)} \left( 2\pi \ri \lambda^{(c)} \right)^{2-2g}, \qquad g\ge 2,
\ee
and we would expect the appropriate instanton action in this frame to be
\be
A^{(c)}_c(\kappa)=-4 \pi^2\lambda^{(c)}. 
\ee
Indeed, one can verify that this is just the analytic continuation of (\ref{wiac}) to $\kappa=-4 \ri$. Similar considerations apply to the other instanton actions in the 
conifold and strong coupling frame. 

\subsection{Large order behavior}

We now provide some numerical evidence that the actions we have found control indeed the large order 
behavior of the genus expansion. We will only consider the behavior in the 
weak coupling frame, but similar considerations and tests can be made for the other frames. For simplicity of notation, in this subsection we will remove 
the superscript $(w)$ in our expressions. Our numerical analysis is done for the original sequence $F_g$ coming from the matrix model. It can be easily shown that 
the redefinition of the $F_g$s which occurs when we use the type IIA parameters, as explained in (\ref{ssdic}), does not change the leading asymptotics (\ref{asymfg}), 
and it only affects the subleading $1/g$ corrections. 

Generically, the instanton actions we have found are complex, and we will write them as 
\be
A_p (\lambda)= \left|A_p(\lambda)\right| \re^{\ri \theta_p(\lambda)}.
\ee
If the genus $g$ amplitudes are real (as it happens for example for 
$\lambda$ and $k$ real), complex 
instantons governing the large order behavior must appear in complex conjugate pairs (this was pointed out already in \cite{blgzj}, in ordinary quantum mechanics). 
This means that, for real $\lambda$, the large order behavior of $F_g(\lambda)$ must be given by 
\be
\label{oscil}
\ba
F_g(\lambda) &\sim \Gamma(2g-1) \left\{ C_p(\lambda) \left(A_p(\lambda)\right)^{-2g} +\overline C_p(\lambda)  \left(\overline A_p(\lambda)\right)^{-2g} \right\} \\
&\sim \Gamma(2g-1) \left| A_p(\lambda)\right| ^{-2g} \cos\left( 2g\theta_p(\lambda) +\delta_p(\lambda) \right), \qquad  g\gg 1.
\ea
\ee
In this equation, $C_p(\lambda)$ is the next correction to the asymptotics, which in some simple matrix models 
can be obtained by a one-loop calculation in the background of an instanton \cite{david,mswone}, and 
\be
\delta_p(\lambda)={\rm arg}\left( C_p(\lambda) \right). 
\ee
The choice of instanton action here depends on the value of $\lambda$, as explained above. If both $\left( A_p(\lambda)\right)^2$ and the $F_g(\lambda)$ are real, 
the large genus behavior is given simply by 
\be
F_g(\lambda) \sim \Gamma(2g-1) \left( A_p(\lambda)\right)^{-2g}, \qquad  g\gg 1.
\ee
This is what happens for example for $\lambda$ imaginary and negative, near the conifold point $\lambda_c$. 

\begin{figure}[!ht]
\leavevmode
\begin{center}
\includegraphics[height=4cm]{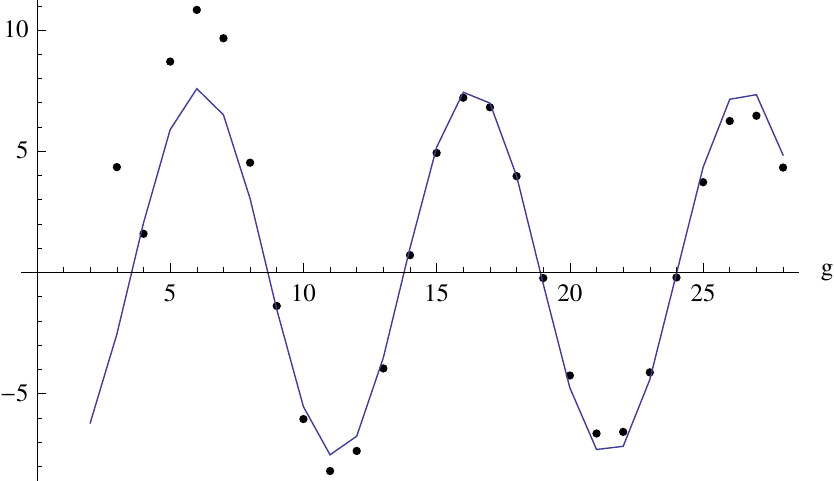}\qquad \qquad 
\includegraphics[height=4cm]{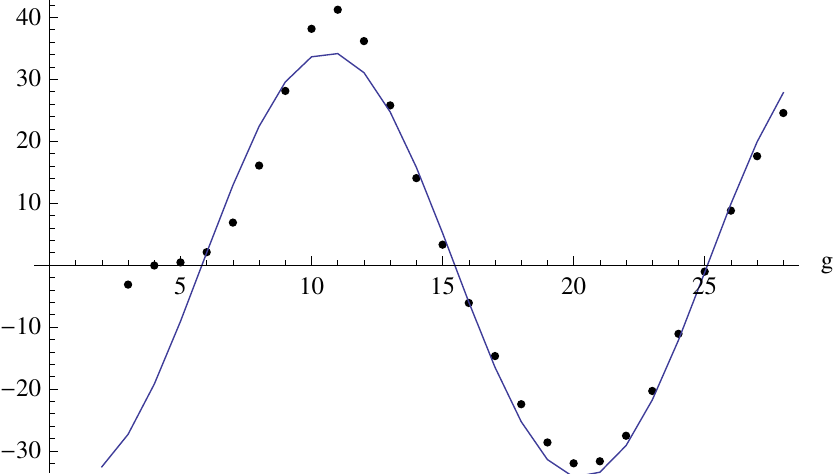}
\end{center}
\caption{In these figures, the dots represent the sequence (\ref{rseq}) for values of $\lambda$ in the strong coupling region: $\lambda\approx 1.2838$ (left) and $\lambda \approx 4.6687$ (right). The action is then $A_s(\lambda)$, given in (\ref{stronga}). The continuous line represents the oscillatory behavior in the r.h.s. of (\ref{cosibe}), where the angle is the one associated to the 
strong coupling action $\theta_s(\lambda)$. 
}
\label{wsactions}
\end{figure} 

When the instanton action is complex, the asymptotics 
is much harder to 
study numerically, since the standard techniques of acceleration of convergence (like Richardson extrapolation) do not apply to the oscillatory behavior 
(\ref{oscil}), and in addition the phase $\delta_p(\lambda)$ is not known. In these cases the 
sequence 
\be
\label{rseq}
R^p_g= (-1)^{g+1} { \pi F_g \over  \Gamma(2g-1) \left| A_p(\lambda)\right| ^{-2g+1}}
\ee
should behave as
\be
\label{cosibe}
R^p_g \sim  \cos\left( 2g\theta_p(\lambda) +g \pi +\delta_p(\lambda) \right), 
\ee
i.e. it should lead to an oscillatory behavior in $g$, with (unknown) constant amplitude but with a known frequency given by $\theta_p(\lambda)$. The factor 
$(-1)^{g+1}$ in (\ref{rseq}) has been introduced for convenience, in view of the forthcoming discussion on Borel summability, and it leads to the shift by $g\pi$ in 
(\ref{cosibe}). 

When the action is real, we can actually extract the value of the instanton action from the sequence 
\be
\label{qseq}
Q^p_g={4 g^2  F_g (\lambda)  \over F_{g+1}(\lambda) \left( A_p(\lambda)\right)^{2}} 
\ee
which as $g\rightarrow \infty$ should asymptote $1$. Standard acceleration methods like Richardson extrapolation can be used to test this behavior to high precision, as in \cite{mswone}.

\begin{figure}[!ht]
\leavevmode
\begin{center}
\includegraphics[height=4cm]{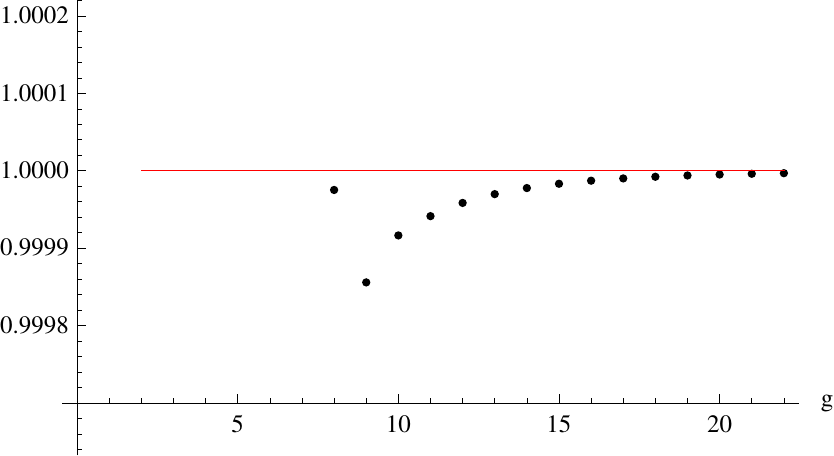}\qquad 
\includegraphics[height=4cm]{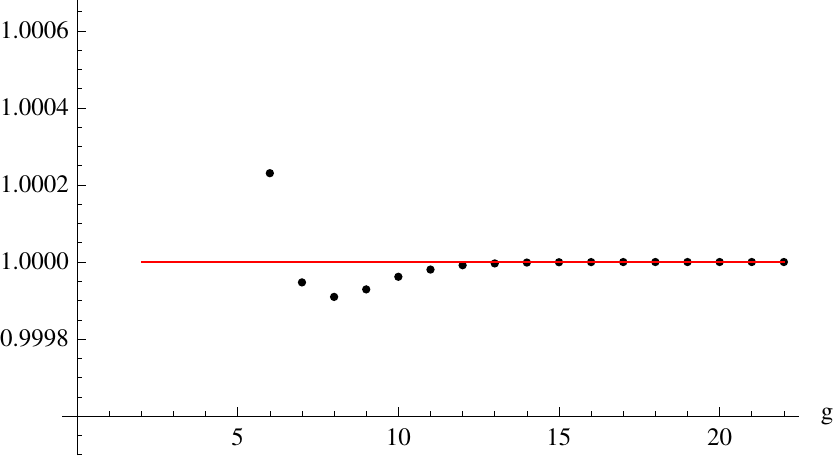}
\end{center}
\caption{In these figures, the dots represent the fifth Richardson transform of the sequence (\ref{qseq}) for values of $\lambda$ along the negative imaginary axis $\lambda\approx -0.1386 \, \ri$ (left) and $\lambda \approx -0.0620 \, \ri$ (right), and for the conifold and the weak coupling action, respectively. They converge quite rapidly to unity, verifying in this way that the proposed instanton actions control the large order behavior of the genus $g$ free energies.}
\label{wcactions}
\end{figure} 

 We now present two tests of the large order behavior of the genus $g$ free energies $F_g$, as predicted by the instanton analysis of the previous subsection. 
 
 For $\lambda$ real and large, we expect the large order behavior to be controlled by the action (\ref{stronga}). This action is complex, and should lead to an oscillatory behavior in $F_g$. 
 We can then compare the sequence $R^s_g$, for $g=2, \cdots, 28$, as computed numerically in (\ref{rseq}), to the expected behavior (\ref{cosibe}). This is done in \figref{wsactions} for two values of $\lambda$ in the 
 strong coupling region. The agreement is rather good. In order to plot the continuous line in these figures, we have taken $\delta_s(\lambda)=
 -2\theta_s (\lambda)$, which leads to a good matching.
 
 When $\lambda$ (or $\kappa$) is on the negative imaginary axis, the relevant instanton actions are the conifold 
 and the weak coupling actions, as shown in the figure on the right in \figref{weakactions}. These are real and pure imaginary, respectively. Therefore, we can use the sequence (\ref{qseq}) and its Richardson transforms to test the expected large order behavior. In this region there is a competition between the conifold and weak coupling instanton actions, and we should pass from a regime dominated by the weak coupling action near $\lambda=0$, to a regime dominated by the conifold action near $\lambda=\lambda_c$. This is precisely what the numerical analysis shows. As an example, we show in \figref{wcactions} the fifth Richardson transform of the sequence (\ref{qseq}) for two different values of $\lambda$ and two different instanton actions: on the left, we consider $\lambda\approx -0.1386 \, \ri$ and the conifold action, while on the right we consider $\lambda \approx -0.0620\,  \ri$ and the 
 weak coupling action. As we see, the expected asymptotic value (unity) is reached quite accurately.

\subsection{Borel summability}

\FIGURE[ht]{
\includegraphics[height=6cm]{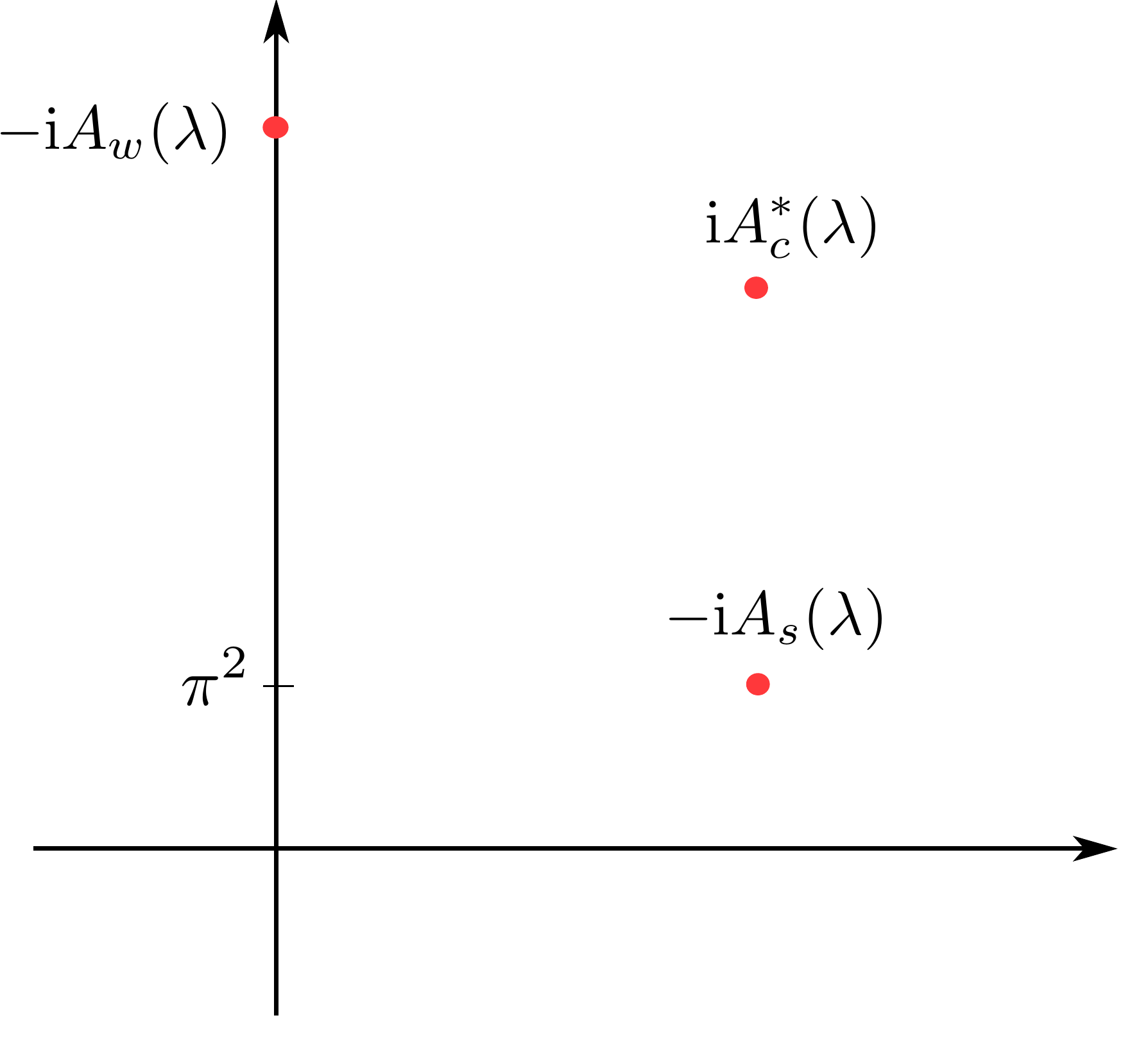} 
\caption{Singularities in the first quadrant of the Borel plane, for $\lambda \sim 1$.} 
\label{borelplane}
}

In the physical ABJM theory, $\lambda$ is real and $g_s$ is purely imaginary. The expansion (\ref{fgs}) should be written in terms of the real coupling constant 
$2\pi/k$, i.e. as
\be
F(\lambda, k)=\sum_{g=0}^{\infty} \left( {2\pi \over k}\right)^{g-2} (-1)^{g-1} F_g(\lambda). 
\ee
We get an extra $(-1)^{g-1}$ sign at each genus, and this is what motivated the introduction of this sign in (\ref{rseq}).
 Equivalently, this leads to an extra $-\ri$ factor in the instanton actions computed above. We can now ask whether this factorially divergent 
 series is Borel summable or not. At strong coupling, the behavior of the genus $g$ free energy $(-1)^{g-1} F_g(\lambda)$ 
 is oscillatory for generic $\lambda$. This is because the strong coupling action (\ref{stronga}), which controls the asymptotics in this regime, is complex: 
 \be
 {\rm Im}\left(  -\ri A_s(\lambda)\right) =\pi^2. 
 \ee
In fact, for large $\lambda$ we have, 
\be
\label{iaction}
-\ri A_s(\lambda)=2 \pi^2 {\sqrt{2 \lambda}} + \pi^2 \ri + \CO\left(\re^{-2 \pi {\sqrt{2\lambda}}}\right), \qquad \lambda \gg 1. 
\ee
This suggests that the $1/N$ expansion is Borel summable for generic values of $\lambda$ in the strong coupling region, as it happens in simple quantum-mechanical examples \cite{blgzj}. 
More precisely, Borel summability requires that there are no instantons with 
positive real action, i.e. that there are no singularities along the positive real axis in the Borel plane of the coupling constant $2\pi/k$. For $\lambda>1/4$ none of the 
actions $A_p(\lambda)$ lies on the positive real axis. This is illustrated in \figref{borelplane}, which shows the singularities in the first quadrant of the Borel plane for $\lambda\sim 1$. They are associated to the instanton actions $-\ri A_s(\lambda)$, $-\ri A_w(\lambda)$, and to the conjugate action $\left( -\ri A_c(\lambda)\right)^*$. Notice that there are other singularities in the other quadrants, related to the ones shown in \figref{borelplane} by flipping the sign and by conjugation. 

Since  
\be
{\rm Im}\left(-\ri A_c(\kappa)\right)=\pi^2 \left(1 -4 \lambda(\kappa)\right), 
\ee
the conifold action is real for $\lambda=1/4$, and we have in principle an obstruction to Borel summability. It might happen that there are other instantons in the theory which we have not identified and lead to singularities in the real axis, even at strong coupling. However, if all the instantons are integer linear combinations of (\ref{wca}) and (\ref{wiac}), as we conjectured before, there will be only a countable set of values of $\lambda$ for which this happens. 

Notice that, at large $\lambda$, the imaginary part of the dominant instanton action is subleading, and the action is approximately real. 
This peculiar behavior, in which Borel summability is lost in some limit 
of the parameter space of the model, has been found before in much more conventional models. Indeed, as shown in \cite{hb}, the classical $O(N)$ one-dimensional spin chain has a Borel summable $1/N$ expansion, where each term in the expansion is itself a function of the temperature. However, in the low temperature limit the imaginary part of the instanton action is subleading. Correspondingly, when each term in the $1/N$ series is truncated to its low-temperature limit, the resulting 
expansion is not Borel summable anymore. Nevertheless, it should be kept in mind that in general the 
asymptotics does not commute with taking limits in parameter space. In the case of ABJM, for example, 
the asymptotics of the strongly coupled $F_g$ (where we neglect worldsheet instanton corrections) is not governed by the strong coupling limit of the instanton action (\ref{iaction}).

We conclude that for generic, real values of $\lambda$ in the strong coupling region, the genus expansion is very likely 
Borel summable. This means that all the information 
about the partition function of the dual superstring theory can be retrieved from the perturbative genus expansion, by Borel resummation. This is in contrast with the 
genus expansion of unitary models coupled to gravity, which is not Borel summable \cite{ezj}. 
Of course, the lack of Borel summability is not a problem if we have an unambiguous 
non-perturbative definition, as it is the case here. It just means that we have to add explicit non-perturbative effects in the theory in a careful way, as illustrated for example 
in \cite{mmnp}. But it is interesting that type IIA superstring on this AdS$_4$ background leads to a Borel resummable string expansion, since this means that it represents a stable background with respect to quantum-mechanical tunnelling effects in the string coupling constant. 

Let us now consider the analytic continuation of the ABJM theory to an {\it imaginary} value of the Chern--Simons level, 
\be
k=\ri \gamma, \qquad \gamma \in \IR, 
\ee
so that $\lambda$ is also imaginary. In this case $g_s$ is real, and the action controlling large order behavior near $\lambda=\lambda_c$ is the conifold 
action, which is also real. The free energies $F_g$ have all the same sign now and the expansion in $1/\gamma$ is {\it not} Borel summable.

\subsection{Double scaling limit near the conifold point}
One interesting aspect of the ABJM free energy is the existence of the critical point (\ref{critpoint}) for an imaginary value of the coupling, which 
corresponds in the Calabi--Yau language to a conifold point. The genus $g$ 
free energies $F_g(\lambda)$ are singular at this point, and their behavior near $\lambda=\lambda_c$ is given by 
\be
F_g\sim C_g \left(2 \pi^2 \ri \left(\lambda-\lambda_c \right) \over \log(\lambda-\lambda_c) \right)^{2-2g}, \qquad g\ge 2,
\ee
where
\be
C_g={B_{2g} \over 2g(2g-2)}.
\ee
This is of course the critical behavior associated to the $c=1$ string at the self-dual radius 
(see for example \cite{klebanov}). The scaling variable is 
\be
\mu \sim {\lambda-\lambda_c  \over \log(\lambda-\lambda_c)}.
\ee
This $c=1$ behavior is expected from the Calabi--Yau point of view \cite{gvafa}, but it is more  
surprising from the point of view of ABJM theory and its string dual. 

The scenario one finds for ABJM theory near 
$\lambda_c$ (i.e. a non-trivial critical point at imaginary 't Hooft coupling, a non-trivial double-scaling limit, and the lack of Borel summability 
which signals an instability) has been advocated in \cite{polyakov} in order to analytically continue AdS theories to de Sitter space. It would be 
interesting to understand this better. 

\section{Instanton at strong coupling}

Based on the AdS/CFT correspondence, we would expect that the instanton configurations that we have found in the matrix model/gauge theory context 
should 
correspond to instanton configurations in the string theory dual. A natural source for such instanton effects are Euclidean D-branes wrapped around submanifolds in 
the target AdS$_4\times \IC\IP^3$. In this section, we want to interpret the strong coupling instanton $A_s$ (which is the dominant configuration in the strongly 
coupled region) as a D-brane configuration, and we will find a D2-brane whose action coincides with the action $A_s$ at large $\lambda$. Notice that, after including 
the coupling constant, the action (\ref{iaction}) becomes, at strong coupling, 
\be
\label{Astring}
{k \over 2 \pi \ri} A_s(\lambda) \approx k \pi {\sqrt{2 \lambda}} + { \pi \ri k \over 2}. 
\ee
In terms of the string coupling constant this can be also written as $A_{\rm st}/g_{\rm st}$, where 
\be
\label{Laction}
A_{\rm st}\approx {1 \over 4} \left(\frac{L}{\ell_s}\right)^3\left(1 + 2\pi \ri{ \ell_s^2\over L^2} \right). 
\ee
The leading part of this action has the appropriate form for a extended object in three dimensions, and it is natural to identify it with an Euclidean D2 brane. In fact, 
it can be written as 
\be
T_{{\rm D2}}{\rm vol}(\IR\IP^3)
\ee
and seems to correspond to an Euclidean D2 brane wrapping a $\IR\IP^3$ inside $\IC\IP^3$. We will now make this more precise by an explicit calculation. Note 
from (\ref{Laction}) that the imaginary part of the instanton action, which makes the $1/N$ expansion Borel summable, is in fact an $\alpha'$ correction. 
This means that Borel summability is in this case a stringy effect, and it is invisible in the supergravity limit.

\subsection{D2-brane instantons}
\label{sec:D2}

We work in the coordinate system for $\IC\IP^3$ given in Appendix~\ref{app:metric}. The metric has the form (\ref{CP3-metric}). We 
consider a D2-brane wrapping the submanifold of fixed $\alpha$ 
with $\vartheta_1=\vartheta_2=\vartheta$ and $\varphi_1=-\varphi_2=\varphi$. The metric is that 
of a warped $\IR\IP^3$ (note that the period of $\chi$ is $2\pi$)
\be
\rd s^2=
\frac{L^3}{4k}\Big[\rd\vartheta^2+\sin^2\vartheta\,\rd\varphi^2
+\sin^2\alpha\,(\rd\chi+\cos\vartheta\,\rd\varphi)^2\Big]\,,
\label{warped}
\ee
and in addition in the world-volume we include a field strength 
$F_{\vartheta\varphi}=E\sin\vartheta$.

The classical action including the Dirac-Born-Infeld (DBI) and Chern--Simons (CS) terms is
\be
\CS_\text{D2}=T_\text{D2}\int \re^{-\Phi}\sqrt{\det(g+2\pi\alpha'F)}
+T_\text{D2}\int\pi \ri\alpha'P[C_1]\wedge F\,,
\label{D2-action}
\ee
where $P[C_1]$ is the pullback to the world-volume of the one-form (\ref{C1}), which 
in our subspace is
\be
C_1=\frac{k}{2}\Big[\cos\alpha(\rd\chi+\cos\vartheta\,\rd\varphi)-\rd\chi-\rd\varphi\Big].
\ee
The extra $\rd\chi$ and $\rd\varphi$ terms, which are exact, make the expression regular 
at $\alpha=\vartheta=0$. Similar terms with opposite signs will be regular at 
$\alpha,\vartheta=\pi$.

Plugging our ansatz in we find
\be
\CS_\text{D2}=\frac{T_\text{D2}L^3}{8}\int \rd\chi\,\rd\vartheta\,\rd\varphi\,\sin\vartheta
\left[\sin\alpha\sqrt{1+\beta^2E^2}
+\ri\beta E(\cos\alpha-1)\right],
\ee
with $\beta=8\pi k/L^3=\sqrt{2/\lambda}$ (setting $\alpha'=1$). Note that we are using conventions where the D2-brane tension is
$T_\text{D2}=1/4\pi^2$.

The equation of motion for $\alpha$ gives the relation
\be
\beta E=-\ri\cos\alpha\,.
\ee
Then the electric flux density is the conserved momentum dual to the electric field
\be
p=\ri\,\frac{\delta\CL}{\delta E}=\beta\sin\vartheta\,.
\ee
The classical action should be expressed in terms of $p$, rather than $E$, and the Legendre transform 
gives
\be
\CS_\text{D2}^\text{classical}
=\frac{T_\text{D2}L^3}{8}\int \rd\chi\,\rd\vartheta\,\rd\varphi\,\ri pE+\CS_\text{D2}=\frac{L^3}{4}=\pi k\sqrt{2\lambda}\,,
\ee
In precise agreement with the leading, real part of the action of the matrix model instanton (\ref{Astring}). 
This is the same as the action of $k/2$ string instantons.

Above we wrote the DBI action suppressing fluctuations in the three orthogonal directions in $\IC\IP^3$. 
It is easy to include them and one finds that this D2-brane is a classical solution which is unstable to 
fluctuations in these three directions.

\subsection{Supersymmetry}

The D-brane we have found should be a BPS state since $\IR\IP^3$ is a generalized Lagrangian submanifold 
in $\IC\IP^3$ (this has been established in a related context in \cite{hk}). We will now confirm this by direct calculation, see \cite{takayanagi,Koerber:2007xk,koerberlectures} for similar considerations.

Our choice of frame and the corresponding expression for the Killing spinors are given 
in Appendix~\ref{app:killing}. For our ansatz (with all the other fields set to zero) they 
are
\be
\epsilon=e^{\frac{\alpha}{4}(\hat\gamma\gamma_4-\gamma_{7\natural})}
e^{\frac{\vartheta}{4}(\hat\gamma\gamma_5-\gamma_{8\natural}+\gamma_{79}+\gamma_{46})}
e^{-\frac{\xi_1}{2}(\hat\gamma\gamma_\natural-\gamma_{47})
-\frac{\xi_2}{2}(\gamma_{58}-\gamma_{69})}
\epsilon_0
=\CM\epsilon_0\,,
\label{resks}
\ee
$\epsilon_0$ is a constant 32-component spinor and the Dirac matrices 
satisfy $\gamma_{012345678 9\natural}=1$.

The angles $\xi_i$ are the phases from (\ref{zi}) 
\be
\xi_1=\frac{\chi+\varphi}{2}\,,\qquad
\xi_2=\frac{\chi-\varphi}{2}\,.
\ee

The supersymmetries preserved by a D2-brane are determined by solving
the following equation on the D2-brane solution
\be\label{projD2}
\Gamma\, \epsilon = \epsilon\,,
\ee
where $\Gamma$ for our D2-brane solution is given by (see e.g.~\cite{Koerber:2007xk})
\be
\Gamma=\frac{\ri}{\CL_{DBI}}\left(\Gamma^{(3)}
+2\pi\alpha'F_{\vartheta\varphi}\, \Gamma^{(1)}\gamma_\natural\right)\, .
\label{GammaD2}
\ee
Here
\be
\Gamma^{(3)}=\Gamma_{\mu_1\mu_2\mu_3}\, 
\frac{\partial x^{\mu_1}}{\partial \sigma^1}
\frac{\partial x^{\mu_2}}{\partial \sigma^2}\,
\frac{\partial x^{\mu_3}}{\partial \sigma^3}\,,
\qquad
\Gamma^{(1)}=\Gamma_\chi\,,
\ee
are the pullback of the curved space-time Dirac matrices in the world-volume 
directions (with and without the directions of the field strength $F_{\vartheta\varphi}$). 
Plugging in our choice of coordinates and the details of the solution we find
\be
\begin{split}
&\Gamma^{(3)} =\frac{1}{8}\sin\alpha\sin\vartheta\,\gamma_{758}\,
e^{\frac{\alpha}{2}(\gamma_{56}-\gamma_{89})}
\,,\\
&2\pi\alpha'F_{\vartheta\varphi}\Gamma^{(1)}
=-\frac{\ri}{8}\cos\alpha\sin\alpha\sin\vartheta\,\gamma_{7}\,,\\
&{\cal L}_{DBI} = \frac{1}{8}\sin^2\alpha\sin\vartheta\,.
\end{split}
\ee
And we therefore find that (\ref{projD2}) reads
\be\label{zwei}
\Big(\ri\,\gamma_{758}\,e^{\frac{\alpha}{2}(\gamma_{56}-\gamma_{89})} 
+\cos\alpha\,\gamma_{7\natural}\Big)\,\epsilon
=\sin\alpha\,\epsilon\,.
\ee
Simple manipulations allow to write this equation as
\be
-\ri\,\gamma_{58\natural}\,e^{\frac{\alpha}{2}(2\gamma_{7\natural}+\gamma_{56}-\gamma_{89})} \,\epsilon
=\epsilon\,.
\ee

Next we need to commute this operator through $\CM$ in (\ref{resks}). As it turns out, 
only the $\alpha$ dependent term in $\CM$ does not commute through and we find the equation
\be
-\ri\,\CM^{-1}\gamma_{58\natural}\,e^{\frac{\alpha}{2}(2\gamma_{7\natural}+\gamma_{56}-\gamma_{89})} \,\epsilon
=-\ri\,\gamma_{58\natural}\,
e^{\frac{\alpha}{2}(\hat\gamma\gamma_4+\gamma_{7\natural}+\gamma_{56}-\gamma_{89})} \,\epsilon_0
=\epsilon_0\,.
\ee
It is easy to see that the operator appearing in this equation squares to unity, and half its 
eigenvalues are $+1$ and half $-1$. Since it does not commute with the $S_i$ operators 
in (\ref{ss}), the D2-brane is $1/2$ BPS.

Note in particular that for $\alpha=0$ we find the equation 
$\ri\,\gamma_{58\natural}\epsilon_0=\epsilon_0$, which is the projector equation for a 
fundamental string wrapping the $\vartheta_1,\varphi_1$ sphere. In this limit the D2-brane 
instanton indeed degenerates to $k/2$ regular string instantons.
While the supercharges at different values of $\alpha$ are not the same, it is possible to choose the 
orientation of the D2-branes in $\IC\IP^3$ such that their 
supersymmetry is compatible with the supercharge used for localization and  with world-sheet 
instantons (of certain orientation). Therefore, these D2-branes 
have the right structure to be responsible for the non-perturbative effects we have found in the matrix model.

\sectiono{Conclusions and open problems}

In this paper we have built on the recent progress in calculating the superstring theory partition function of ABJM theory at all genera \cite{dmp} in order 
to study nonperturbative effects in the string coupling constant. Conceptually, our approach is similar to what had been pursued in the 
case of noncritical strings \cite{shenker,polchinski, martinec,akk}: first, we extract spacetime instanton actions from the large genus behavior, and 
then we try to identify some of these instantons with Euclidean D-brane configurations. Of course, our setting is more complex since 
all relevant quantities are non-trivial functions of the 't Hooft parameter. In the calculation of the superstring free energies we have used the large $N$ dual description of ABJM theory in terms of a matrix model \cite{kapustin}, and we 
have formulated a general strategy to extract spacetime instantons from matrix models which generalizes previous results in, for example, \cite{mswone,kmr}. 

Our work raises many questions and can be pursued in various ways. First of all, since our description of matrix model instantons is made in the language of 
special geometry, it should determine the large order behavior of 
the genus $g$ amplitudes in general topological string models, not necessarily encoded in matrix integrals. It would be interesting to study simple models with 
a spectral curve description, like topological string theory on local $\IP^2$, in order to test the method and learn about possible non-perturbative 
structures in topological strings. The general picture we have developed might shed further light in related contexts, like the models studied in \cite{cdv}. 

It would be very interesting as well to obtain a deeper understanding of the ABJM matrix model instantons. One should study in more 
detail the D2-brane interpretation of the instanton that dominates at strong coupling. In particular, it would be interesting 
to identify the source of the imaginary, $\alpha'$-correction to the instanton action in (\ref{Laction}). Another interesting issue is the interpretation of the 
process of eigenvalue tunneling. As pointed in \cite{mp,dt}, 
the ABJM matrix model is closely related to that of Chern--Simons theory on $\IS^3/\IZ_2$. 
The non-perturbative partition function of Chern-Simons requires summing over all partitions $(N_1, N_2)$ of 
the total numbers of eigenvalues $N$ between the two cuts \cite{mmnp,mpp}. For ABJM one performs an analytical continuation such 
that the total number of eigenvalues is zero, i.e. one considers the sector with eigenvalues $(N, -N)$. Then one has regular ``particle'' eigenvalues in one of the cuts 
while the other cut has ``hole'' eigenvalues. Moving eigenvalues 
between the cuts, which in Chern--Simons on $\IS^3/\IZ_2$ implements the sum over partitions, changes in ABJM theory 
the total rank of both gauge groups, since it leads to the process 
\be
(N, -N) \rightarrow (N\pm 1, -N\mp 1). 
\ee
%
One interesting question is whether this process, which is ultimately responsible for the appearance of the 
strong coupling instanton, has an interpretation directly in the gauge theory or in the superstring dual. 
Finally, it would be important to identify the weakly coupled instanton (\ref{wca}) as a D-brane configuration. It is easy to see that 
in string units this action scales like $L^4$, so it seems to correspond to an Euclidean D4-brane, but since its action is purely imaginary 
in ABJM theory it is not obvious what is its geometric meaning.

\section*{Acknowledgments}
We would like to thank Pepe Barb\'on, Rajesh Gopakumar, 
Stefan Hohenegger, Daniel Jafferis, Ingo Kirsch, Paul Koerber, Juan Maldacena, Ruben Minasian, Greg Moore, 
Ricardo Schiappa and Alireza Tavanfar for 
useful conversations. N.D. is grateful for the hospitality of the University of Geneva Mathematics 
Department in the final stages of this project. The work of N.D. is underwritten by an advanced fellowship of the 
Science \& Technology Facilities Council. 
The work of M.M. and P.P. is supported in part by the Fonds National Suisse. P.P. is also supported by FASI RF 14.740.11.0347.


\appendix

\section{Metric}
\label{app:metric}

In this appendix we follow the notations in \cite{dp} (with the replacement $\chi\to2\chi$). 
The metric on AdS$_4\times\IC\IP^3$ is
\be
\rd s^2=\frac{L^3}{4k}\left(\rd s^2_{\text{AdS}_4}+4\rd s^2_{\IC\IP^3}\right)\,.
\label{metric}
\ee
For the AdS$_4$ part we may use the global Lorentzian metric
\be
\rd s_{\text{AdS}_4}^2=-\cosh^2\rho\,\rd t^2+\rd\rho^2
+\sinh^2\rho\big(\rd\theta^2+\sin^2\theta\,\rd\psi^2\big)\,.
\label{AdS-metric}
\ee

The metric on $\IC\IP^3$ can be written in terms of four complex projective
coordinates $z_i$ as
\be
\rd s_{\IC\IP^3}^2=\frac{1}{\rho^2}\sum_{i=1}^4 \rd z_i\,\rd\bar z_i
-\frac{1}{\rho^4}\bigg|\sum_{i=1}^4 z_i\,\rd\bar z_i\bigg|^2\,,\qquad
\rho^2=\sum_{i=1}^4|z_i|^2\,.
\ee

In the following we choose a specific representations in terms of angular 
coordinates (used also in \cite{Cvetic:2000yp,Nishioka:2008gz}). 
We start by parametrizing $\IS^7\subset\IC^4$ as
\be
\begin{aligned}
z_1&=\cos\frac{\alpha}{2}\cos\frac{\vartheta_1}{2}\,\re^{\ri(2\varphi_1+2\chi+\zeta)/4}\,,
&\qquad
z_3&=\sin\frac{\alpha}{2}\cos\frac{\vartheta_2}{2}\,\re^{\ri(2\varphi_2-2\chi+\zeta)/4}\,,\\
z_2&=\cos\frac{\alpha}{2}\sin\frac{\vartheta_1}{2}\,\re^{\ri(-2\varphi_1+2\chi+\zeta)/4}\,,
&\qquad
z_4&=\sin\frac{\alpha}{2}\sin\frac{\vartheta_2}{2}\,\re^{\ri(-2\varphi_2-2\chi+\zeta)/4}.
\end{aligned}
\label{zi}
\ee

The metric on $\IS^7$ is then given by
\begin{align}
\rd s^2_{\IS^7}=&\frac{1}{4}\Bigg[
\rd\alpha^2
+\cos^2\frac{\alpha}{2}(\rd\vartheta_1^2+\sin^2\vartheta_1\,\rd\varphi_1^2)
+\sin^2\frac{\alpha}{2}(\rd\vartheta_2^2+\sin^2\vartheta_2\,\rd\varphi_2^2)
\nonumber\\&\hskip1cm
+\sin^2\frac{\alpha}{2}\cos^2\frac{\alpha}{2}
(2\rd\chi+\cos\vartheta_1\,\rd\varphi_1-\cos\vartheta_2\,\rd\varphi_2)^2
+\frac{1}{4}(\rd\zeta+2A)^2\, \Bigg]\,,
\label{S7-metric}
\\
A=&\cos\alpha\,\rd\chi+\cos^2\frac{\alpha}{2}\cos\vartheta_1\,\rd\varphi_1
+\sin^2\frac{\alpha}{2}\cos\vartheta_2\,\rd\varphi_2\,.
\label{S7-form}
\end{align}
The angle $\zeta$ appears only in the last term and if we drop it 
we end up with the metric on $\IC\IP^3$
\be
\begin{aligned}
\rd s^2_{\IC\IP^3}=\frac{1}{4}\bigg[&
\rd\alpha^2
+\cos^2\frac{\alpha}{2}(\rd\vartheta_1^2+\sin^2\vartheta_1\,\rd\varphi_1^2)
+\sin^2\frac{\alpha}{2}(\rd\vartheta_2^2+\sin^2\vartheta_2\,\rd\varphi_2^2)
\\&
+\sin^2\frac{\alpha}{2}\cos^2\frac{\alpha}{2}
(2\rd\chi+\cos\vartheta_1\,\rd\varphi_1-\cos\vartheta_2\,\rd\varphi_2)^2\bigg].
\end{aligned}
\label{CP3-metric}
\ee
The ranges of the angles are
$0\leq\alpha,\vartheta_1,\vartheta_2\leq\pi$ and
$0\leq\varphi_1,\varphi_2,\chi\leq2\pi$.

In addition to the metric, the supergravity background has the
dilaton, and the 2-form and 4-form field strengths from the
Ramond-Ramond (RR) sector
\be
\re^{2\Phi}=\frac{L^3}{k^3}\,,
\qquad
F_4=\frac{3}{8}\,L^3\,\rd\Omega_{\text{AdS}_4}\,,
\qquad
F_2=\frac{k}{2}\,\rd A\,.
\label{field-strengths}
\ee
Here $\rd\Omega_{\text{AdS}_4}$ is the volume form on AdS$_4$ and $F_2$ is
proportional to the K\"ahler form on $\IC\IP^3$.

To write down the general D-brane action in this background one also 
needs the potentials for these forms. 
The one-form potential is, up to gauge transformations
\be
C_1=\frac{k}{2}\,A\,,
\label{C1}
\ee
with $A$ defined in (\ref{S7-form}). It is easy to write down $C_3$, 
the three-form potential for $F_4$ and $C_5$, its magnetic dual, but they are not 
required for our calculation in Section~\ref{sec:D2}.

The relation between the parameters of the string background and of
the field theory are (for $\alpha'=1$ and in the supergravity and tree-level limit)
\be
\frac{L^3}{4k}=\pi\sqrt{\frac{2N}{k}}=\pi\sqrt{2\lambda}\,.
\label{radius-1}
\ee

\section{Killing spinors}
\label{app:killing}

To write down the Killing spinors it is useful to start in 11-dimensions with 
the AdS$_4$ metric in (\ref{AdS-metric}) and the $\IS^7$ metric in (\ref{S7-metric}).

We take the elfbeine (ignoring the factor of $L^3/k$)
\be
\begin{gathered}
e^0=\frac{1}{2}\cosh\rho\,\rd t\,,\quad
e^1=\frac{1}{2}\,\rd\rho\,,\quad
e^2=\frac{1}{2}\sinh\rho\,\rd\theta\,,\quad
e^3=\frac{1}{2}\sinh\rho\sin\theta\,\rd\psi\,,\\
e^4 =\frac{1}{2}  \rd\alpha, \qquad 
e^5 =\frac{1}{2}  \cos\frac{\alpha}{2}\,\rd\vartheta_1,\qquad 
e^6 =\frac{1}{2}  \sin\frac{\alpha}{2} \, \rd\vartheta_2,\\
e^7 = \frac{1}{2}  \cos\frac{\alpha}{2}\sin\frac{\alpha}{2} \Bigl( \cos \vartheta_1 \,
\rd\varphi_1 - \cos\vartheta_2\,\rd\varphi_2 + 2\rd\chi \Bigr),\\
e^8 = \frac{1}{2}  \cos\frac{\alpha}{2} \sin\vartheta_1\,\rd\varphi_1,\quad
e^9 =\frac{1}{2}  \sin\frac{\alpha}{2}\sin\vartheta_2 \, \rd\varphi_2\,,\\
e^\natural=-\frac{1}{4}\left(\rd\zeta+2\cos^2\frac{\alpha}{2}\cos\vartheta_1\,\rd\varphi_1
+2\sin^2\frac{\alpha}{2}\cos\vartheta_2\,\rd\varphi_2+2\cos\alpha\,\rd\chi\right).
\end{gathered}
\ee

Killing spinor equation for this background comes from 
the supersymmetry transformation of the gravitino
\be
\delta\Psi_\mu=D_\mu\epsilon-\frac{1}{288}\Big(
\Gamma_\mu^{\,\nu\lambda\rho\sigma}
-8\delta_\mu^\nu\Gamma^{\lambda\rho\sigma}\Big)F_{\nu\lambda\rho\sigma}\epsilon\,,\qquad
D_\mu\epsilon=\partial_\mu\epsilon+\frac{1}{4}\omega_\mu^{ab}\gamma_{ab}\epsilon\,.
\ee
The 4-form corresponding to the AdS$_4 \times \IS^7$ solution is
$F_{\nu\lambda\rho\sigma} = 6\, \varepsilon_{\nu\lambda\rho\sigma}$, 
where the epsilon symbol is the volume form on AdS$_4$ (so the indices take 
the values $0,1,2,3$). Plugging this into the variation
above one finds the Killing spinor equation
\be
\begin{aligned}
D_\mu \epsilon&=\hat \gamma\Gamma_\mu \epsilon\,,\qquad& \mu&=0,1,2,3\\
D_\mu \epsilon&=\frac{1}{2} \hat \gamma\Gamma_\mu \epsilon\,,\qquad& \mu&=4,5,\cdots,9,10
\end{aligned}
\ee
where $\mu$ runs over all 11 coordinates, and $\hat \gamma=\gamma^{0123}$. Note that 
small $\gamma$ have tangent-space indices while capital $\Gamma$ carry 
curved-space indices.

The general solution to these equations is
\be
\re^{\frac{\alpha}{4} ( \hat \gamma\gamma_4 - \gamma_{7\natural}   ) }
\re^{\frac{\vartheta_1}{4} (  \hat \gamma\gamma_5 - \gamma_{8\natural}  ) }
\re^{\frac{\vartheta_2}{4} ( \gamma_{79} + \gamma_{46} ) }
\re^{-\frac{\xi_1}{2} \hat \gamma\gamma_\natural}
\re^{-\frac{\xi_2}{2} \gamma_{58}}
\re^{-\frac{\xi_3}{2} \gamma_{47}}
\re^{-\frac{\xi_4}{2} \gamma_{69}}
\re^{\frac{\rho}{2}\hat\gamma\gamma_1}
\re^{\frac{t}{2}\hat\gamma\gamma_0}
\re^{\frac{\theta}{2}\gamma_{12}}
\re^{\frac{\psi}{2}\gamma_{23}}\epsilon_0
={\mathcal M}\epsilon_0\,,
\label{Killing}
\ee
where the $\xi_i$ are given by
\be
\xi_1=\frac{2\varphi_1+2\chi+\zeta}{4}\,,\qquad
\xi_2=\frac{-2\varphi_1+2\chi+\zeta}{4}\,,\qquad
\xi_3=\frac{2\varphi_2-2\chi+\zeta}{4}\,,\qquad
\xi_4=\frac{-2\varphi_2-2\chi+\zeta}{4}\,.
\ee
In (\ref{Killing}) $\epsilon_0$ is a constant 32-component spinor 
and the Dirac matrices were chosen such that
$\gamma_{012345678 9\natural}=1$. 
A similar calculation in a different coordinate system was done in 
\cite{Nishioka:2008ib}.

To see which Killing spinors survive the orbifolding from M-theory to type IIA, we 
write the spinor $\epsilon_0$ in a basis which diagonalizes
\be
\ri\hat\gamma\gamma_\natural\epsilon_0=s_1\epsilon_0\,,\qquad
\ri\gamma_{58}\epsilon_0=s_2\epsilon_0\,,\qquad
\ri\gamma_{47}\epsilon_0=s_3\epsilon_0\,,\qquad
\ri\gamma_{69}\epsilon_0=s_4\epsilon_0\,.
\label{ss}
\ee
All the $s_i$ take values $\pm1$ and by our conventions on the product 
of all the Dirac matrices, the number of negative eigenvalues is even. 
Now consider a shift along the $\zeta$ circle, which changes all the 
angles by $\xi_i\to\xi_i+\delta/4$, the Killing spinors transform 
as
\be
{\mathcal M}\epsilon_0\to {\mathcal M} \re^{\ri\frac{\delta}{8}
(s_1+s_2+s_3+s_4)}\epsilon_0\,.
\ee
This transformation is a symmetry of the Killing spinor when two of the $s_i$ eigenvalues 
are positive and two negative and not when they all have the same sign (unless $\delta$ 
is an integer multiple of $4\pi$). Note that on $\IS^7$ the radius of the $\zeta$ circle 
is $8\pi$, so the $\IZ_k$ orbifold of $\IS^7$ is given by taking $\delta=8\pi/k$. 
The allowed values of the $s_i$ are therefore
\be
(s_1,s_2,s_3,s_4)\in\left\{
\begin{matrix}
(+,+,-,-),\ (+,-,+,-),\ (+,-,-,+),\\
(-,+,+,-),\ (-,+,-,+),\ (-,-,+,+)
\end{matrix}
\right\}
\label{signs}
\ee
Each configuration represents four supercharges, so the orbifolding 
breaks $1/4$ of the supercharges (except for $k=1,2$) and leaves 
24 unbroken supersymmetries.

\end{document}